\documentclass[%
 reprint,
 showkeys,
superscriptaddress,
amsmath,amssymb,
aps,
pra,
]{revtex4-2}

\usepackage{float}
\usepackage{mathrsfs}
\usepackage{graphicx}
\usepackage{physics}
\usepackage{hyperref}
\usepackage{cleveref}
\usepackage{bbm}
\usepackage{booktabs}%
\usepackage[dvipsnames]{xcolor}
\hypersetup{
        colorlinks   = true,
        citecolor    = LimeGreen,
        linkcolor    = LimeGreen}

\begin{document}

\preprint{APS/123-QED}

\title{MEMO-QCD: Quantum Density Estimation through \\ Memetic Optimisation for Quantum Circuit Design}

\author{Juan E. Ardila-García}
\email{juardilag@unal.edu.co}
\affiliation{Grupo de Superconductividad y Nanotecnología, Departamento de Física, Universidad Nacional de Colombia, Bogotá, 111321, Colombia}

\author{Vladimir Vargas-Calderón}
\email{vladimir.vargas@zapata.ai}
\affiliation{Zapata AI, 100 Federal St., Boston, 02110, MA, USA}

\author{Fabio A. González}
\affiliation{MindLab, Departamento de Ingenieria de Sistemas e Industrial, Universidad Nacional de Colombia, Bogotá, 111321, Colombia}

\author{Diego H. Useche}
\affiliation{MindLab, Departamento de Ingenieria de Sistemas e Industrial, Universidad Nacional de Colombia, Bogotá, 111321, Colombia}

\author{Herbert Vinck-Posada}
\affiliation{Grupo de Superconductividad y Nanotecnología, Departamento de Física, Universidad Nacional de Colombia, Bogotá, 111321, Colombia}

\date{\today}

\begin{abstract}
This paper presents a strategy for efficient quantum circuit design for density estimation. The strategy is based on a quantum-inspired algorithm for density estimation and a circuit optimisation routine based on memetic algorithms. The model maps a training dataset to a quantum state represented by a density matrix through a quantum feature map. This training state encodes the probability distribution of the dataset in a quantum state, such that the density of a new sample can be estimated by projecting its corresponding quantum state onto the training state. We propose the application of a memetic algorithm to find the architecture and parameters of a variational quantum circuit that implements the quantum feature map, along with a variational learning strategy to prepare the training state. Demonstrations of the proposed strategy show an accurate approximation of the Gaussian kernel density estimation method through shallow quantum circuits illustrating the feasibility of the algorithm for near-term quantum hardware.
\end{abstract}

                   
\maketitle

\section{Introduction}\label{sec1}

Developing efficient and scalable methods for probability distribution estimation, or density estimation, is a central task in the field of machine learning \cite{Bigdeli2020LearningGM, Nachman2020AnomalyDW, Hu2020AnomalyDU, Fraley2002ModelBasedCD, Bortoloti2020SupervisedKD}.
Under the assumption that data is drawn from an underlying probability distribution, many methods strive for learning this distribution. This pursuit has led to the development of classical algorithms that can be divided into (i) methods that fit observed data to a previously assumed probability distribution $p(\boldsymbol{x}|\boldsymbol{\theta})$ which depends on a number of parameters that are optimised according to the data, i.e. parametric algorithms \cite{Vapnik1999SupportVM, Papamakarios2017MaskedAF, Varanasi1989ParametricGG}, and (ii) methods that do not assume any particular probability distribution, having the ability to reproduce arbitrary probability distributions from data, i.e. non-parametric algorithms \cite{Wang2019NonparametricDE, Parzen1962OnEO, Rosenblatt1956RemarksOS}. Recently, quantum algorithms have been also explored as alternatives for performing density estimation, including parametric \cite{Liang2019QuantumAD,  Guo2021QuantumAF, Verdon2019QuantumHM, PhysRevA.108.062422, Schuhmacher2023UnravellingPB} and non-parametric methods \cite{Gonzlez2020ClassificationWQ, Useche2021QuantumMC, Gonzalez2021LearningWD, nakayama2024explicitquantumsurrogatesquantum}, where one can find quantum kernel methods and Hamiltonian learning methods. 

Typically, a model excels in capturing an underlying probability distribution when it is biased towards the intended distribution--which is, in principle, unknown.
However, in scenarios with large number of data, the kernel density estimation \cite{Rosenblatt1956RemarksOS, Parzen1962OnEO} (KDE) method reliably converges to the underlying probability distribution due to the law of large numbers.
Hence, KDE stands out as one of the main approaches for estimating unknown probability distributions.
One of the downsides of KDE is that inference time scales with the size of the training data.
Recently, González et al. \cite{Gonzalez2021LearningWD} proposed a quantum-inspired algorithm for probability density estimation--dubbed density matrix kernel density estimation (DMKDE)--whose inference time does not depend on the training sample size, yet provides a reliable approximation of KDE.

At the core of DMKDE is the utilisation of mixed states as mathematical objects to describe probability distributions.
These states, inherently normalised, provide a natural foundation for capturing intricate data patterns.
Roughly speaking, DMKDE consists on mapping every training sample into the quantum state of a system, i.e. a data point $\vb*{x}$ is mapped to $\ketbra{\psi(\vb*{x})}$.
Then, the average over the states corresponding to training data points forms a classical superposition of quantum states $\rho = \frac{1}{N}\sum_{i=1}^N\ketbra{\psi(\vb*{x}_i)}$.
Finally, to estimate the density at a particular point in space, this point is mapped in the same way as one did with training samples to get a new quantum state, i.e., $\vb*{x}_\star\mapsto \ketbra{\psi(\vb*{x}_\star)}$; all that is needed to estimate the probability density at this point is to project the mixed quantum training state onto this new quantum state, i.e. $\expval{\rho}{\psi(\vb*{x}_\star)}$, and this overlap is the density estimation, up to a normalisation constant that can be determined.
One can appreciate that only an expected value is needed to estimate the density, and that this operation is independent of the size of the training dataset once the training state $\rho$ is known.

The DMKDE method has been implemented as a quantum-inspired machine learning method that runs on classical computers, achieving good results in a variety of tasks such as density estimation for anomaly detection, classification, regression, among others~\citep{Gonzlez2020ClassificationWQ, Useche2021QuantumMC, Gonzalez2021LearningWD, GallegoMejia2022InQMADIQ, GallegoMejia2022FastKD, GallegoMejia2022QuantumAF, GallegoMejia2022LEANDMKDEQL, BustosBrinez2022ADDMKDEAD, ToledoCorts2022GradingDR}.
Being a quantum-inspired technique, it is natural to ask whether this method can be implemented in quantum hardware. 

In Ref.~\citep{VargasCalderon2022OptimisationfreeDE}, an implementation of this method on quantum circuits was proposed, where the quantum state representing the training data set was a pure state, i.e. a superposition state $\ket{\Psi} = \frac{1}{N}\sum_{i=1}^N\ket{\psi(\vb*{x}_i)}$, instead of a classical mixture of quantum states.
Estimation of the probability density at a point $\vb*{x}_\star$ was achieved by estimating $\abs{\braket{\Psi}{\psi(\vb*{x}_\star)}}^2 = \abs*{\bra{0}U^\dagger_{\mathscr{D}}U_\star\ket{0}}^2$, where $\ket{0}$ is the zero product state of a multi-qubit register, $U_{\mathscr{D}}$ is a unitary that prepares the training state and $U_\star$ is a unitary that prepares the state corresponding to the point $\vb*{x}_\star$.
However, this implementation contains a bottleneck which prevents its scalability, being that the unitary that prepares the state of each new data point must be found through an arbitrary state preparation algorithm, representing an exponentially complex problem in the number of qubits.
In addition, the unitary that prepares the training state must also be found and compiled, and although it must be done only once for each training dataset, it is also an exponentially complex problem that prevents scalability.

In this work, we propose an implementation of DMKDE in quantum circuits, where the training state is represented not as a pure state (cf. Ref.~\citep{VargasCalderon2022OptimisationfreeDE}) but as a quantum mixed state, faithful to the original proposal in Ref.~\citep{Gonzalez2021LearningWD}.
To do this, we define the training state unitary matrix $U_{\mathscr{D}}$ on an extended qubit system that comprises ancillary qubits, where we trace out their degrees of freedom to end up with a proper quantum mixed state.
We show how to find this circuit, as well as the unitary that prepares the state of new samples $U_{\star}$.
To overcome the scalability problems identified in Ref.~\cite{VargasCalderon2022OptimisationfreeDE}, firstly, we propose using a memetic algorithm~\cite{Moscato1999MemeticAA}--which combines variational quantum circuit architecture search using a genetic algorithm~\citep{AltaresLpez2021AutomaticDO} with the optimisation of the variational quantum circuit parameters through regular stochastic gradient descent \cite{Ruder2016AnOO}--to find a unique unitary that approximately prepares the state of any new sample. Secondly, we propose a method to find--in a variational way--the unitary that prepares the training state by means of the optimisation of a variational quantum circuit~\cite{Cerezo2020VariationalQA} with a hardware efficient ansatz~\cite{Kandala2017HardwareefficientVQ} (HEA). Our proposal addresses the aforementioned scalability bottlenecks of Ref.~\citep{VargasCalderon2022OptimisationfreeDE}, enabling the implementation of DMKDE on current hardware~\cite{Preskill2018QuantumCI, Brandhofer2021SpecialSN}.

It is noteworthy that \citet{Useche2022QuantumDE} implemented a non-parametric quantum-classical density matrix density estimation algorithm based on DMKDE, named \mbox{Q-DEMDE} by introducing a quantum extension of random and adaptive Fourier features, along with a quantum algorithm for estimating the expected value of a density matrix from its eigendecomposition.
However, similar challenges as in Ref.~\citep{VargasCalderon2022OptimisationfreeDE} are present in \mbox{Q-DEMDE}--namely, exact state preparation and explicit density matrix decomposition are needed.

The structure of the article is as follows: in \cref{sec:dmkde} we present an overview of the DMKDE algorithm.
In \cref{sec:dmkde_on_circuit} we show how DMKDE can be implemented in quantum circuits.
In \cref{sec:QED} we propose to use a memetic algorithm to variationally find $U_{\star}$ and show how to optimise a variational quantum circuit of fixed-architecture to find $U_{\mathscr{D}}$.
In \cref{sec:results} we characterise the variational preparation of both the training state and the states of new samples, and we showcase the implementation of DMKDE~\cite{Gonzalez2021LearningWD} by means of these variational quantum circuits and compare our results with \mbox{Q-DEMDE}~\cite{Useche2022QuantumDE}.
Then, in \cref{sec:discussion} we discuss the contributions, advantages and limitations of our proposal.
Finally, in \cref{sec:conclusions} we establish our conclusions and future outlook.

\section{Preliminary on the DMKDE algorithm}
\label{sec:dmkde}

The problem of probability density estimation is: given a data set $\{\vb*{x}_i\}_{i=1}^N$ assumed to have been drawn from a probability distribution $f$, come up with a function $\hat{f}$ such that the distance between $f$ and $\hat{f}$ is minimised, for some specified distance.
Of course, one does not have access to $f$, so the distance between it and $\hat{f}$ is to be estimated by means of data.
Density matrix kernel estimation (DMKDE)~\cite{Gonzalez2021LearningWD, Gonzlez2020ClassificationWQ} works as an efficient approximation of kernel density estimation~\citep{Parzen1962OnEO, Rosenblatt1956RemarksOS}, which is a non-parametric method that solves such task.
The DMKDE algorithm is defined by the following steps:

\begin{enumerate}
    \item Quantum feature mapping: each training data point $\boldsymbol{x}_i\in\mathbb{R}^d$ is mapped to $\ket{\psi(\boldsymbol{x}_i)}$ using a quantum feature map (QFM)
    \begin{equation} 
        \psi: \mathbb{R}^d\to \mathcal{H},
        \label{eq:1}
    \end{equation}
    that maps data from $\mathbb{R}^d$ to quantum states that belong to a Hilbert space associated to some quantum system.
    We choose a separable QFM in the data features, meaning that
    \begin{equation}
        \ket{\psi(\boldsymbol{x}_i)}=\bigotimes_{j=1}^d\ket*{\varphi(x_i^j)},
        \label{eq:2}
    \end{equation}
    where $x_i^j$ refers to the $j$-th component of the $i$-th data point of the training set.
    The single-feature QFM $\varphi$ thus maps real numbers to the single-feature Hilbert space, denoted by $\mathcal{H}_x$.
    It follows that $\mathcal{H} = \mathcal{H}_x^{\otimes d}$.

    The QFM $\psi$ induces a quantum kernel defined by
    \begin{align}
        \abs{k(\vb*{x}, \vb*{x}')}^2 = \abs{\bra{\psi(\vb*{x})}\ket{\psi(\vb*{x}')}}^2.\label{eq:3}
    \end{align}
    An important remark is that the induced kernel is isotropic thanks to the QFM $\psi$ being separable in the data features.
    
    \item Training state construction~\footnote{This is called a training state because building it resembles the usual training step in machine learning methods. However, in principle, there does not need to be an actual training procedure to build such training state.}: a classical superposition of quantum states is constructed from all the $N$ samples of the training set in the form
    \begin{equation}
        \rho_\text{train}=\dfrac{1}{N}\sum_{i=1}^N \ketbra{\psi(\boldsymbol{x}_i)}.
        \label{eq:4}
    \end{equation}
    Note that this state is no longer separable in the data features.

    \item Projection: given a new sample $\boldsymbol{x_{\star}}\in\mathbb{R}^d$, its associated state is constructed with the QFM $\psi$.
    As explained in Ref.~\cite{Gonzlez2020ClassificationWQ}, the density estimation consists of projecting the training state onto the state of the new sample:
    \begin{equation}
        \hat{f}(\boldsymbol{x_{\star}})\propto\langle \psi(\boldsymbol{x}_{\star}) |\rho_\text{train} |\psi(\boldsymbol{x_{\star}} )\rangle = \frac{1}{N}\sum_{i=1}^N \abs*{k(\vb*{x}_\star,\vb*{x}_i)}^2.
        \label{eq:5}
    \end{equation}
\end{enumerate}

\section{DMKDE on Quantum Circuits}
\label{sec:dmkde_on_circuit}

In this section, we propose how to use quantum circuits as the platform for performing density estimation via mixed states. 
We consider a circuit of $dn_x + n_a$ qubits, where $n_x$ qubits are used to encode a single feature, $d$ is the number of features, and $n_a$ is the number of auxiliary qubits needed to prepare the mixed training state (see~\cref{eq:4}).
To keep a consistent notation, from now on, we will write states of multi-qubit systems in the decimal representation.
Additionally, we will index kets by `f', `d' or `a' to indicate that they are elements of the Hilbert space corresponding to single-feature quantum states, data point quantum states, or quantum states in the auxiliary system, respectively.
For example, for $x\in\mathbb{R}$, its quantum feature mapped state is $\ket{\varphi(x)}_\text{f} = \sum_{k=1}^{2^{n_x}}\braket{\varphi(x)}{k}_\text{f}\ket{k}_\text{f}\in\mathcal{H}_x$.
Similarly, for a data point $\vb*{x}\in\mathbb{R}^d$, its quantum feature mapped state is $\ket{\psi(\vb*{x})}_\text{d} = \bigotimes_{j=1}^d\ket*{\varphi(x^j)}_{\text{f}}\in\mathcal{H}$.
States in the auxiliary system will be written in the basis $\{\ket{k}_\text{a}\}_{k=1}^{2^{n_a}}$, where $\ket{k}_\text{a}\in\mathcal{H}_a$ for any $k$.
States belonging to the bipartite space $\mathcal{H}\otimes\mathcal{H}_a$ will not have any subscript.
The following steps define the complete protocol for density matrix kernel density estimation on a qubit-based quantum computer.

\begin{enumerate}
    \item Training state circuit: given a dataset $\mathscr{D}$ of $N$ data points in $\mathbb{R}^d$, we define the training state circuit $U_{\mathscr{D}}$ as an $n=dn_x+n_a$-qubit circuit that satisfies
    \begin{align}
        \Tr_a(U_\mathscr{D}\ketbra{0}U^\dagger_{\mathscr{D}}) = \rho_\text{train},\label{eq:partial_trace_Ud}
    \end{align}
    where the trace is over the $n_a$ auxiliary qubit degrees of freedom.
    Such circuit can be easily constructed if one has access to the probability amplitudes of the training data point quantum states, i.e., $\ket{\psi(\vb*{x}_i)}_\text{d}=\sum_\alpha c_{i}^\alpha\ket{\alpha}_\text{d}$.
    In the expression for the probability amplitude $c_i^\alpha$ the subscript indexes the element of the training data set, whereas the superscript indexes the component of the state vector.
    In appendix~\ref{app:training_state_unitary} we show how to exactly build this circuit.
    For this, one needs at most $dn_x$ auxiliary qubits for exactly representing the state $\rho_\text{train}$.
    In this work, we build it by training a variational quantum circuit (see~\cref{sec:training_state_circuit}).
    This results in the circuit $U_\mathscr{D}$, depicted in~\cref{fig:density_circuit}.

\begin{figure}[b]
    \centering
    \includegraphics{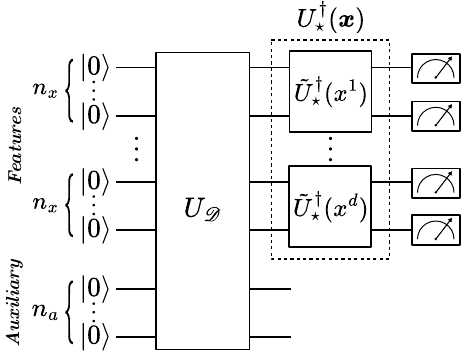}
    \caption{Circuit to implement density estimation. $U_{\mathscr{D}}$  is the unitary that prepares the training state and $U_{\star}(\boldsymbol{x}_{\star})=\bigotimes_{j=1}^d \tilde{U}_{\star}(x^j_\star)$ is the unitary that prepares the state of the new sample. Density estimation is done by projecting the state of a new sample $|\psi(\boldsymbol{x}_{\star})\rangle $ onto the training state $\rho_\text{train}$, which is equivalent to measuring the bit string $\boldsymbol{0{}}$ in this circuit, ignoring auxiliary qubits, as explained in the main text.}
    \label{fig:density_circuit}
\end{figure}

    \item New sample circuit: a quantum circuit $U_{\star}$ prepares the state $\ket{\psi(\boldsymbol{x}_{\star})}_\text{d}$ of a new sample $\boldsymbol{x}_{\star}$.
    A way to exactly prepare the desired quantum state is to represent the circuit by any unitary matrix $U$ that satisfies $U^\alpha_0 = c_\star^\alpha$, where $\ket{\psi(\vb*{x}_\star)}_\text{d} = \sum_\alpha c_\star^\alpha \ket{\alpha}_\text{d}$.
    This fixes a row of the matrix $U$, and the rest of the rows can be obtained using the Gram-Schmidt procedure.
    In this work, we combine memetic algorithms and variational quantum circuits to find the representation of the circuit $U_\star$ (see~\cref{sec:QFM_circuit}), which is depicted in~\cref{fig:density_circuit}.

\begin{figure*}
    \centering
    \includegraphics[scale = 1]{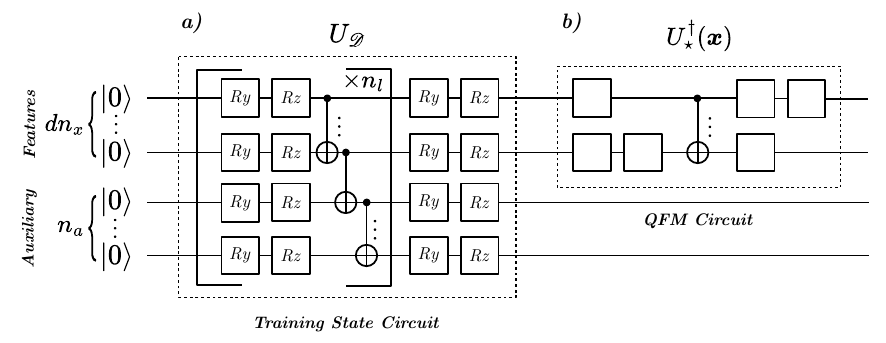}
    \caption{DMKDE quantum circuit synthesised using the Quantum Circuit Design (QCD) method. (a) Training circuit: is composed of a HEA architecture comprising the QFM and auxiliary qubits. The first block, composed of $R_y, R_z$ gates, and a cascade of CNOT gates is repeated a number $n_l$ of times (number of layers), followed by a final block of $R_y$ and $R_z$ gates. (b) Quantum Feature Map (QFM) circuit: the architecture, derived from a memetic algorithm (see~\cref{sec:QFM_circuit} for details), approximates the Gaussian kernel taking into account to the number of qubits used for feature encoding and the desired depth.}
    \label{fig:DMKDE_circuit}
\end{figure*}
 
    \item Estimation: The probability density
    \begin{equation}
        \hat{f}(\boldsymbol{x}_{\star})\propto\bra{0}_\text{d} U_{\star}^{\dagger} \Tr_\text{a}(U_\mathscr{D}\ketbra{0}U^\dagger_{\mathscr{D}})U_{\star}\ket{0}_\text{d}
    \label{eq:8}
    \end{equation}
    can be estimated using a quantum circuit (see \cref{fig:density_circuit}), by making $M$ measurements on the non-auxiliary qubits (effectively carrying out the partial trace), so that $\hat{f}(\boldsymbol{x}_{\star})\propto M_{\boldsymbol{0}}/M$, where $M_{\boldsymbol{0}}$ is the number of times that the $\boldsymbol{0}$ bit string is measured. 

\end{enumerate}

Kernels are central to this work because our density estimation model performance will ultimately be linked to how well we can use quantum circuits to approximate such kernels (cf.~\cref{eq:3}).
In particular, DMKDE approximates the widely used KDE with a Gaussian kernel.
We highlight the fact that one could use any other kernel.
It is important to mention that the quantum states constructed by the QFM can correspond to infinite dimensional states, and in fact, it is known that coherent states induce the Gaussian kernel \citep{Gonzlez2020ClassificationWQ}.
However, in a quantum circuit it is only possible to represent states in finite Hilbert spaces.
Thus, the kernel induced by a QFM that maps data to the space of quantum qubits can be an approximation of the theoretically expected kernel. 

\section{Quantum Circuit Design (QCD)}
\label{sec:QED}

The previous section proposed a general way to implement DMKDE in quantum circuits.
We showed the conditions that the unitaries that realise the circuits in~\cref{fig:density_circuit} must satisfy (see appendix~\ref{app:training_state_unitary}).
The decomposition of these unitaries into one- and two-qubit gates is left for compilation routines, which can be computationally costly.
One of the main bottlenecks of this approach is that compilation of a circuit is needed for every single new data point at which we want to estimate the probability density.
To overcome these issues, in this section we propose methods to approximate the unitaries in~\cref{fig:density_circuit} that prepare the training state $\rho_\text{train}$ and the state of a new sample $\ket{\psi(\boldsymbol{x}_{\star})}_\text{d}$.
Our proposal seeks to make a DMKDE implementation on quantum circuits that is scalable and applicable to real problems in the current NISQ computer era, where we aim to address primarily the depth limitation of quantum circuit hardware \cite{Preskill2018QuantumCI, Brandhofer2021SpecialSN}.

The construction of the DMKDE circuit is divided into two parts: first, a circuit (QFM circuit) that approximates a desired kernel is found; then, the circuit that prepares the mixed training state is constructed.
The whole circuit construction is based on the correct approximation of the kernel.
Broadly speaking, this is accomplished by exploring the space of trained variational ansätze with a memetic algorithm.
This memetic algorithm is a modified version of the genetic algorithm in Ref.~\citep{AltaresLpez2021AutomaticDO}, with a small improvement (see \ref{sec:QFM_circuit}), to propose variational ansätze that are optimised through gradient descent techniques~\citep{Cerezo2020VariationalQA}.
As per the circuit that prepares the training state, good results were obtained by just using a shallow hardware efficient ansatz (HEA) architecture~\citep{Kandala2017HardwareefficientVQ}, which is optimised by means of a global cost function through gradient descent techniques.
Details for the training state preparation are given in~\cref{sec:training_state_circuit}. The complete architecture for synthesising the circuit that implements DMKDE is shown in \cref{fig:DMKDE_circuit}.

\subsection{Quantum Feature Map Circuit}
\label{sec:QFM_circuit}

As discussed previously, kernels can be induced by QFMs that map data to quantum states.
In this paper, we choose the Gaussian kernel with bandwidth $\gamma$, defined by
\begin{align}
    \abs{k(\vb*{x}, \vb*{x}')}^2 = \exp(-\gamma\norm{\vb*{x} - \vb*{x}'}_2),
\end{align}
where $\norm{\cdot}_2$ is the $L_2$ norm, because it is the most widely used kernel for density estimation.
A QFM $\psi$ that induces the Gaussian kernel is the coherent state QFM~\citep{Gonzlez2020ClassificationWQ}, which maps data to the mean number of photons of such states.
Of course, the Hilbert space to which coherent states belong is infinite dimensional, so mapping coherent states to quantum circuits requires truncating this Hilbert space basis.
Our aim, however, is not to approximate the coherent quantum states, but to come up with quantum states defined on qubit systems that approximately induce the kernel; a subtle, but important difference.
In other words, we want to find a unitary matrix $U_\star(\vb*{x})$ acting on $dn_x$ qubits such that

\begin{align}
\begin{aligned}
\abs{\bra{0}_\text{d}U^\dagger_\star(\vb*{x})U_\star(\vb*{x}')\ket{0}_\text{d}}^2 \propto& \exp(-\gamma\norm{\vb*{x} - \vb*{x}'}_2)\\
&\quad= \prod_{i=1}^de^{-\gamma(x_i-x_i')^2},
\end{aligned}
\end{align}
where the isotropicity of the Gaussian kernel has been made explicit.
This motivates the chosen separability of the QFM (cf.~\cref{eq:2}), which implies that $U_{\star}(\boldsymbol{x})$ is also separable:
\begin{equation}
    U_{\star}(\boldsymbol{x})=\bigotimes_{j=1}^d \tilde{U}_{\star}(x^j),
    \label{eq:9}
\end{equation}
significantly simplifying the task to solve.
Namely, we no longer need to find $U_\star(\vb*{x})$ acting on $dn_x$ qubits, but $\tilde{U}_\star(x)$, acting on $n_x$ qubits.

\begin{figure}[]
    \centering
    \includegraphics[scale=1]{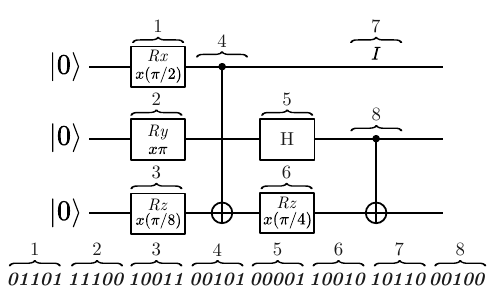}
    \caption{Example of the decoding of a bit string (chromosome) of 8 gates in a quantum circuit (individual). The bit string is divided into sections of 5 bits. Each section encodes a gate. The first gate in the chromosome is applied to the first qubit. The second gate is applied to the second qubit, and so on. When the last qubit is reached, one starts from the first qubit again, until all gates in the chromosome are applied. The specific rules for building a gate from a section of 5 bits are given in the main text and in \cref{tab:bitstrings_left} and \cref{tab:bitstrings_right}.}
    \label{fig:decodification}
\end{figure}

\begin{figure}
    \centering
\includegraphics[scale = 0.9]{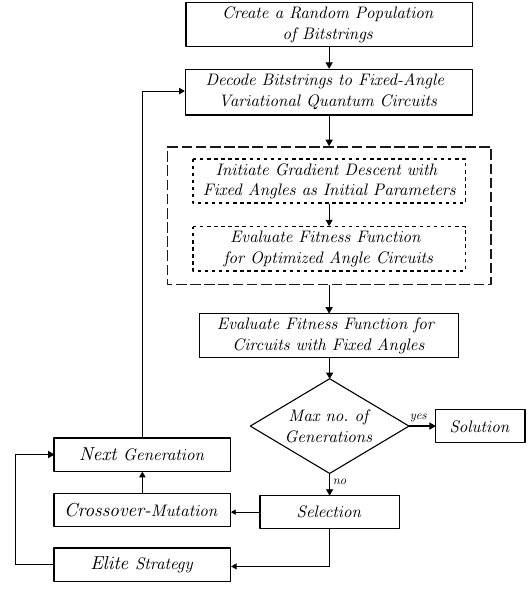}
    \caption{
    Flowchart of the evolutionary optimisation algorithm used to find the best variational quantum circuit that approximates the kernel, and consequently the QFM. The distinction between a genetic and a memetic algorithm lies in the inclusion of local improvers in the latter. In our approach, the local improvement entails optimising the decoded angles of each circuit prior to evaluating their fitness function. This enables the memetic algorithm to explore both the space of variational circuit architectures and parameter space. Boxes with dashed lines denote operations performed when memetic optimisation is selected, utilising \cref{eq:single_feature_cost_fn}. Otherwise, genetic optimisation is conducted without local improvement, employing \cref{eq:genetic_fitness}.}    
    \label{fig:genetic_routine}
\end{figure}

To find such unitary, we parameterise the single-feature circuit $\tilde{U}_\star(x)$ with a bit string $\vb*{b}$ that is decoded into a specific variational ansatz with variational parameters $\vb*{\theta}_{\vb*{b}}$, (see~\cref{fig:decodification} for more details).
It is crucial to understand that the bit string $\vb*{b}$ introduced here has no relation to quantum measurements or computational basis states.
Instead, these bit strings are solely used to encode the structure of the variational ansatz itself, serving as a key element in our parameterisation of $\tilde{U}_\star(x)$.
This distinction is important to avoid any confusion between the encoding scheme and quantum outcomes.
The initial value of the parameters $\vb*{\theta}_{\vb*{b}}$ is also determined by the bit string $\vb*{b}$, as explained in \cref{fig:genetic_routine}.
Therefore, the problem of finding $\tilde{U}_\star(x)$ can be casted as the following optimisation problem:

\begin{widetext}
\vspace*{5px}
\begin{equation}
    \min_{\vb*{b},\vb*{\theta}_{\vb*{b}}}\sum_{(x,x')\sim U^2(a,b)}\left(|k(x,x')|^2  - \abs{\bra{0}_\text{f}\tilde{U}^\dagger_\star(\vb*{b},\vb*{\theta}_{\vb*{b}},x)\tilde{U}_\star(\vb*{b},\vb*{\theta}_{\vb*{b}},x')\ket{0}_\text{f}}^2\right)^2,
    \label{eq:single_feature_cost_fn}
\end{equation}
\vspace*{5px}
\end{widetext}
where the sum is over pairs of numbers sampled from the uniform distribution $U^2(a,b)$ over the square $[a,b]\times[a,b]$.
To solve this problem, we use a memetic algorithm.
The memetic algorithm consists of a genetic algorithm that explores the space of ansätze~\citep{AltaresLpez2021AutomaticDO} (i.e., the space of bit strings $\vb*{b}$), and a gradient-descent algorithm that optimises the variational parameters $\vb*{\theta}_{\vb*{b}}$ of each of the explored ansätze. \Cref{fig:genetic_routine} shows the optimisation process of the memetic algorithm. 

\begin{table}
    \centering
    \begin{tabular}{cc}
        \cmidrule[\heavyrulewidth]{1-2}
        First 3 bits & Quantum gate \\
        \cmidrule(lr){1-2}
        $000$ & Hadamard \\
        $011$ & $R_x$ \\
        $111$ & $R_y$ \\
        $100$ & $R_z$ \\
        $001$ & CNOT \\
        Other bits & Identity \\
        \cmidrule[\heavyrulewidth]{1-2}
    \end{tabular}
    \caption{Set of unitaries applied according to the first three bits of the bitstring.}
    \label{tab:bitstrings_left}
\end{table}

A crucial aspect of genetic and memetic algorithms is the chromosomal representation of possible solutions to the problem (a chromosome represents a point in the solution space \cite{Katoch2020ARO}).
Typically, chromosomes are represented as bit-strings with two possible genes: $0$ and $1$. Since our goal is to find the best architecture for a quantum circuit, we need to devise a method for encoding circuits into bit-strings (chromosomes). 
We adopt the approach proposed by \citet{AltaresLpez2021AutomaticDO}, which uses a 5-bit encoding system per quantum gate. In this system, the first three bits determine the type of quantum gate to be applied, as detailed in \cref{tab:bitstrings_left}. The remaining two bits specify the rotation angle when the selected gates are $R_x$, $R_y$, or $R_z$ (\cref{tab:bitstrings_right} (up)).
In the original implementation \cite{AltaresLpez2021AutomaticDO}, only adjacent qubits could be entangled. However, we have extended this capability by using the last two bits to select positions for the CNOT gates as well. The options for these last two bits are presented in \cref{tab:bitstrings_right} (down). An example of how a bitstring (or chromosome) is decoded into a quantum circuit (individual) is shown in \cref{fig:decodification}.

Once the unitary $\tilde{U}\star$ is determined, the complete QFM unitary $U\star$ can be constructed by vertically stacking the scalar unitaries $\tilde{U}_\star$, as shown in~\cref{fig:density_circuit}. In this manner, the QFM circuit is synthesised for any number of features, being one of the two parts in the circuit construction that implements DMKDE, as shown in \cref{fig:DMKDE_circuit} (b).

\begin{table}[H]
    \centering
    \begin{minipage}{.45\linewidth}
        \centering
        \begin{tabular}{cc}
            \cmidrule[\heavyrulewidth]{1-2}
            \multicolumn{2}{c}{Rotation Gates} \\
            Last 2 bits & Rotation Angle \\
            \cmidrule(lr){1-2}
            $00$ & $\pi$ \\
            $01$ & $\pi/2$ \\
            $10$ & $\pi/4$ \\ 
            $11$ & $\pi/8$ \\
            \cmidrule[\heavyrulewidth]{1-2}
        \end{tabular}
        \vspace{1em}
        \begin{tabular}{cc}
            \cmidrule[\heavyrulewidth]{1-2}
            \multicolumn{2}{c}{CNOT Gate} \\
            Last 2 bits & CNOT position \\
            \cmidrule(lr){1-2}
            $00$ & $i+1$ \\
            $01$ & $i+2$ \\
            $10$ & $i+3$ \\
            $11$ & $i+4$ \\
            \cmidrule[\heavyrulewidth]{1-2}
        \end{tabular}
    \end{minipage}
    \caption{Set of angles or positions determined by the last two bits of the bitstring. The top table shows the rotation angles for the gates, while the bottom table indicates the positions of the CNOT gates, where $i$ represents the qubit on which the gate is applied.}
    \label{tab:bitstrings_right}
\end{table}

\subsection{Training State Circuit}
\label{sec:training_state_circuit}

Once the circuit that prepares the QFM is built, the circuit that prepares the training set quantum mixed state can be built.
This is done with a parameterised quantum circuit with a shallow HEA architecture~\footnote{As in the case of finding the QFM circuit, to find the training state circuit we also used genetic and memetic algorithms for optimising the quantum circuit architecture, but got little to no improvement with respect to the shallow HEA architecture (which showed good results) while incurring in a significant computational time overhead.}, that spans over the QFM and auxiliary qubits to prepare a mixed state in the feature qubits, as shown in~\cref{fig:DMKDE_circuit} (a) and ~\cref{fig:density_circuit}.

The optimisation is performed by maximising the log-likelihood of the data points within the dataset $\mathscr{D}=\{\vb*{x}_i\}_{i=1}^N$, i.e.,
\begin{widetext}
\begin{align}
    \max_{\vb*{\theta}} \ \log\left(\sum_{i=1}^N  \bra{0}_\text{d} U_{\star}^{\dagger}(\vb*{x}_i) \Tr_a(U_\mathscr{D}(\vb*{\theta})\ketbra{0}U^\dagger_{\mathscr{D}}(\vb*{\theta}))U_{\star}(\vb*{x}_i)\ket{0}_\text{d}\right).
\end{align}
\end{widetext}
where $U_{\star}(\vb*{x}_i)$ is the QFM unitary found by the memetic algorithm when solving~\cref{eq:single_feature_cost_fn}.

Maximising the log-likelihood of the data points encoded with the circuit ensures that the variational training circuit accurately encodes the dataset, effectively capturing its underlying structure and enabling reliable density estimation for new feature vectors, as explained in~\cref{sec:dmkde_on_circuit}.

\section{Results}
\label{sec:results}

\subsection{Quantum Feature Map Circuit}

\label{sec:qfm_results}

\begin{figure}[!h]
    \centering
    \includegraphics{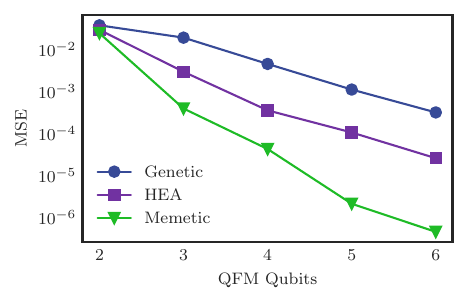}
    \caption{Mean squared error (MSE) of the Gaussian kernel approximation, as a function of the number of qubits $n_x$ that encode a single feature.
    The genetic algorithm (blue circles) optimises the loss by exploring the variational ansatz space with coarse-grained variational parameters (MSE calculated using~\cref{eq:genetic_fitness}).
    Optimisation of the variational parameters of HEA (purple squares) is done through gradient-descent (MSE calculated using~\cref{eq:hea_loss}).
    The memetic algorithm (green triangles) uses gradient-descent to fine-tune the variational parameters of ansätze explored by the genetic algorithm (MSE calculated using~\cref{eq:single_feature_cost_fn}).}    
\label{fig:QFM_problem}
\end{figure}

\begin{figure}[]
    \centering
    \includegraphics[scale = 1]{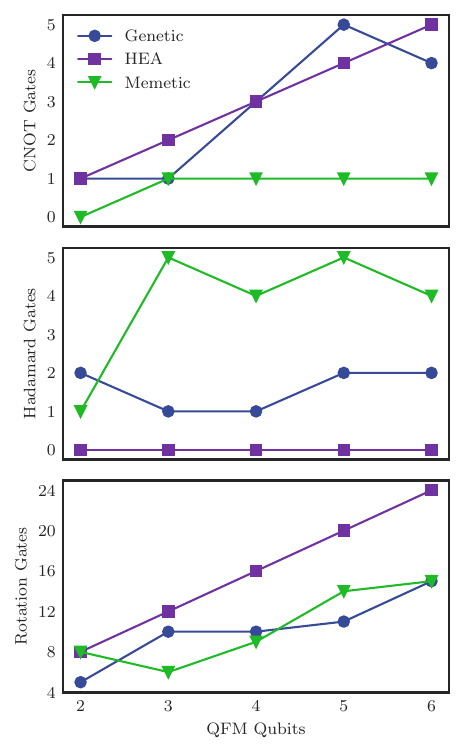}
    \caption{Gate count of variational quantum circuits architectures found by genetic (blue circles) and memetic (green triangles) algorithms compared to the fixed HEA architecture (purple squares). }
    \label{fig:circuits_char}
\end{figure}

To assess the effectiveness of our proposed quantum circuit optimisation, which approximates the kernel using a memetic algorithm, we compared it with both a genetic algorithm and a gradient-descent algorithm employing a fixed hardware-efficient ansatz (HEA) architecture (see appendix \cref{app:fitness_and_loss} for details).
Notably, we constrained the HEA architecture to a single layer to ensure shallow quantum circuits, accommodating current hardware limitations.
Consequently, we also constrained the depth of the genetic and memetic architectures to be less than or equal to that of the HEA architecture.
\Cref{fig:QFM_problem} reports the mean squared error of the Gaussian kernel approximation as a function of the number of qubits for quantum feature-mapping a single feature, showing the superiority of the memetic algorithm.
\Cref{fig:circuits_char} shows a gate count comparison between the best circuits achieved by each of the QFM optimisation methods.
Experimental details associated with all QFM results can be found in \cref{app:experimental_parameters}.

\subsection{Training State Circuit}

\begin{figure*}
    \centering
    \includegraphics[scale = 1.02]{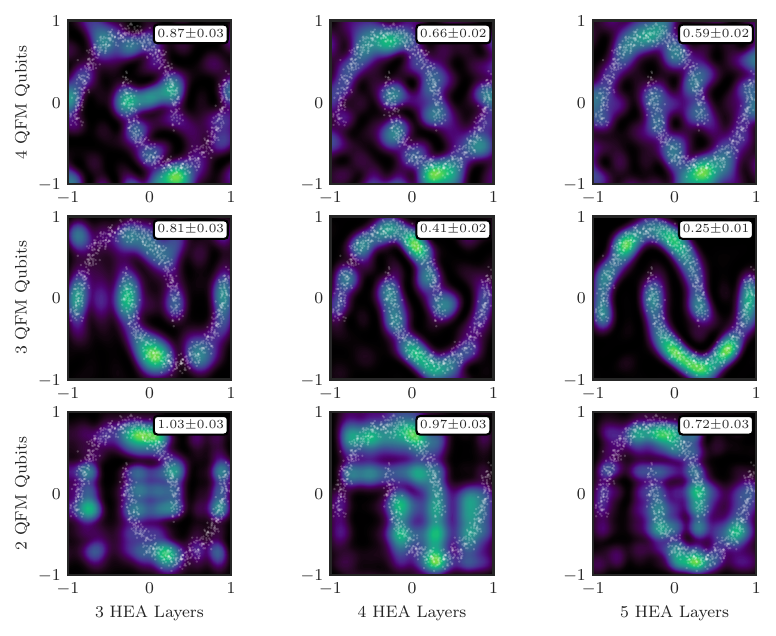}
    \caption{Density maps for a 2D two-moon dataset composed of $1000$ points (shown as small semi-transparent white dots) as a function of the number of qubits of the QFM and the layers of the HEA training state circuit. In this case, two features and one auxiliary qubit are considered, resulting in a total of $2n_x+1$ qubits for the training circuit, where $n_x$ represents the number of QFM qubits. The continuous Kullback-Leibler Divergence (KLD), displayed in the white boxes of each figure, serves as a metric for comparing the density estimates of each model. This metric measures the distance between the estimated probability density and the actual probability density of the training set.}
    \label{fig:memetic-layers-qubits}
\end{figure*}

The construction of the training circuit is based on the circuit found for the QFM~\footnote{Due to its superior performance, only the QFM circuit found by the memetic algorithm is used to build the training circuit, and consequently, the DMKDE circuit.}.
However, it is necessary to determine the number of layers in the training circuit to ensure optimal DMKDE algorithm performance.
We optimise the training circuit using a 2D two-moon dataset (see \cref{fig:memetic-layers-qubits}) consisting of $1000$ points, varying the number of QFM qubits in the range of $[2,4]$ and the number of training circuit layers in the range of $[2,5]$.
We conduct training with $5000$ epochs and a learning rate of $0.4$.
Once the training circuit was determined, we constructed the DMKDE circuit (shown in \cref{fig:density_circuit}) and estimated the density at different points.
\Cref{fig:memetic-layers-qubits} displays density maps for the dataset depending on the number of QFM qubits and training circuit layers.
It also presents the performance of the DMKDE circuit measured through the estimation of the continuous Kullback-Leibler divergence~\citep{PrezCruz2008KullbackLeiblerDE} (KLD).
The KLD serves as a measure of the distance between the density distribution reconstructed from the DMKDE circuit estimation and the density distribution of the dataset (further details are provided in Appendix~\ref{app:data_KLD}).
Finally, \cref{fig:best_model} displays density maps for different datasets estimated by the best DMKDE circuit and reports the KLD estimate.

\begin{figure*}
    \centering
    \includegraphics[scale = 1.02]{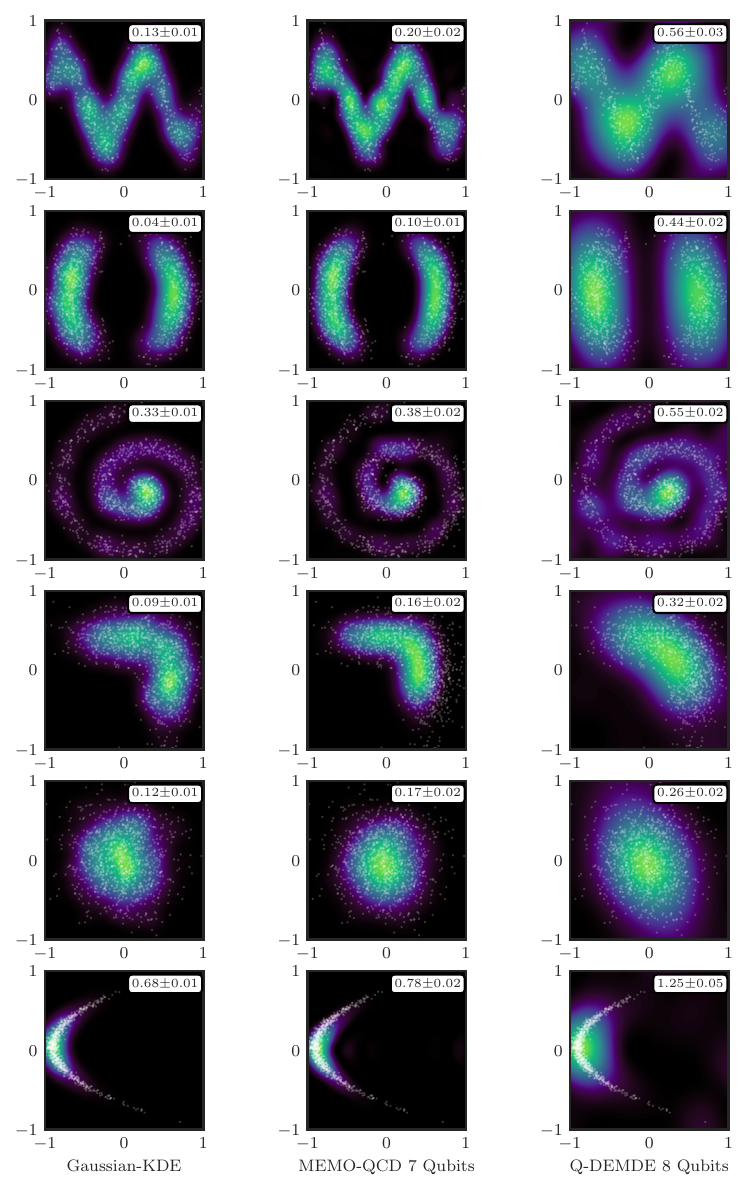}
    \caption{Comparison of MEMO-QCD and Q-DEMDE quantum circuits implementing DMKDE in 2D density estimation with Gaussian KDE for various datasets containing 2000 samples each (data points shown as semi-transparent white dots). The DMKDE circuit employs $3$ qubits in the QFM, $5$ layers in the training circuit, and $1$ auxiliary qubit, leading to a quantum circuit of $7$ qubits. On the other hand, the Q-DEMDE circuit uses 16 random Fourier features (RFF) to encode the features, leading to a quantum circuit of $8$ qubits. The continuous Kullback-Leibler Divergence (KLD), displayed in the white boxes of each figure, serves as a metric for comparing the density estimates of each model. This metric measures the distance between the estimated probability density and the actual probability density of the training set. }
    \label{fig:best_model}
\end{figure*}

\section{Discussion}
\label{sec:discussion}

In this work, we have addressed the problem of estimating the probability density of a data point $\vb*{x}_\star$ given a dataset $\mathscr{D}$, which requires being able to prepare a quantum mixed state $\rho_\text{train}$ for the training data set, and a pure quantum state $\ket{\psi(\vb*{x}_\star)}$ for the new data point.
This proposal, originally envisioned in Ref.~\citep{VargasCalderon2022OptimisationfreeDE}, relied on arbitrary quantum state preparation algorithms, which yield deep circuits not amenable to NISQ devices.
Further, for every single new data point $\vb*{x}_\star$, a new run of such arbitrary quantum state preparation algorithm would be needed.
So far, our work has tackled both of these bottlenecks by the use of variational quantum circuits--found by memetic algorithms--that (i) are shallow, meaning that the resulting circuits are amenable to NISQ devices, and (ii) prepare quantum states for new data points straight-forwardly, eliminating the need for costly arbitrary quantum state preparation algorithms. \Cref{fig:QFM_problem} demonstrates that the memetic algorithm surpasses both a genetic algorithm and a gradient descent algorithm (which optimises circuits of fixed HEA architecture) across different numbers of QFM qubits. 

\Cref{fig:circuits_char} illustrates that architectures discovered by the memetic algorithm, which explores the largest search space, feature minimal entanglement and numerous Hadamard gates.
Conversely, the genetic algorithm yields architectures with higher entanglement but fewer Hadamard gates. 
This indicates that entanglement is less critical for approximating the Gaussian kernel--in line with results from Ref.~\citep{bowles2024better} where it is shown that many quantum ML applications in the literature usually do not require exploiting entanglement--as the memetic algorithm achieves the best performance among the three architectures.
Additionally, the fixed HEA architecture exhibits a linear increase in the number of CNOT gates, similar to the genetic algorithm, but it has a higher count of rotation gates compared to the other two architectures.

We adopt a variational approach to training state preparation, aiming to be compatible with current hardware and to use a single, versatile architecture (HEA) suitable for any dataset. However, using a HEA architecture with a reduced number of layers presents a limitation in our proposal. 
This prevents the circuit from achieving the necessary expressiveness \cite{Nakaji2020ExpressibilityOT, Sim2019ExpressibilityAE} to effectively solve the density estimation problem as the number of qubits increases.
Consequently, the DMKDE circuit fails to scale properly with the number of QFM qubits, despite offering better kernel approximations, as illustrated in \cref{fig:memetic-layers-qubits}.
Nevertheless, we achieve excellent approximations to Gaussian KDE when a low number of layers in the training circuit do not lead to expressiveness issues. 
Specifically, the combination of three qubits in the QFM, five layers in the training circuit, and one auxiliary qubit (a total of seven qubits) yields excellent results for 2D density estimation, as demonstrated in \cref{fig:best_model}, where a direct comparison with KDE is shown. 
Furthermore, \cref{fig:best_model} demonstrates that the 7-qubit configuration MEMO-QCD circuit outperforms the Q-DEMDE circuit, which uses 8 qubits, in achieving a closer approximation to Gaussian KDE. Since MEMO-QCD and Q-DEMDE are specifically designed to approximate KDE, a direct comparison with this method is both relevant and justified.

Despite achieving excellent results for density estimation in two dimensions, scalability issues arise when the number of dimensions increases while maintaining a low number of qubits for each QFM component. 
This is because the number of qubits increases linearly as the number of features grows.
In this case, it may be more effective to bypass the separation of the Gaussian vector kernel into its components. Instead, directly approximating the vector kernel using the QFM circuit can help maintain a fixed number of coding qubits and allow for an increase in the number of layers. 
This approach ensures scalability for higher-dimensional datasets and can be seamlessly implemented with the proposed memetic algorithm, requiring only changes to the encoding of data. 
As shown in \cref{sec:QFM_circuit}, encoding a scalar feature $x^k$ (recall that we are using superscripts to index the component of vectors) involves multiplying it by a parameter $\theta^k$, which defines the angle of a quantum gate as $x^k\theta^k$. 
For a vector feature $\boldsymbol{x}$, it suffices to perform a scalar multiplication with a parameter vector  $\boldsymbol{\theta}_i$ (for the $i$-th parameterised gate) and define the quantum gate angle as $\boldsymbol{x}\cdot\boldsymbol{\theta}_i$ \cite{Woloshyn2024BoseHubbardMW}. Thus, the fitness function for training the vector kernel would take the form

\begin{widetext}
\begin{equation}
    \min_{\vb*{b},\vb*{\theta}_{\vb*{b}}}\sum_{(x,x')\sim U^2(a,b)}\left(|k(\boldsymbol{x},\boldsymbol{x}')|^2  - \abs{\bra{0}_\text{f}U^\dagger_\star(\vb*{b},\vb*{\theta}_{\vb*{b}},\boldsymbol{x})U_\star(\vb*{b},\vb*{\theta}_{\vb*{b}},\boldsymbol{x}')\ket{0}_\text{f}}^2\right)^2.
    \label{eq:vectorial_feature_cost_fn}
\end{equation}
\end{widetext}

\section{Conclusions}
\label{sec:conclusions}

In this work, we implemented a method for performing density estimation in shallow quantum circuits based on the proposals of \citet{Gonzalez2021LearningWD, Gonzlez2020ClassificationWQ}.
We improved the implementation of \citet{VargasCalderon2022OptimisationfreeDE} by representing the training state as a quantum mixed state, ensuring a direct approximation to KDE \cite{Gonzalez2021LearningWD}.
Additionally, we proposed the application of a memetic algorithm to find optimised architectures of variational quantum circuits for preparing the states of new samples and a variational quantum circuit with a fixed HEA architecture for preparing the training state.
Our proposal has enabled us to overcome the bottleneck of preparing new sample states and address the scalability problem in compiling unitaries in quantum circuits, both of which were limitations identified in Ref.~\citep{VargasCalderon2022OptimisationfreeDE}.
Moreover, by taking a different approach from that proposed by \citet{Useche2022QuantumDE} for implementing DMKDE in quantum circuits, we achieved better results in estimating the probability density using fewer qubits and low-depth circuits that are suitable for current quantum hardware. 

The results of the Gaussian kernel approximation demonstrate that the memetic algorithm effectively designs architectures tailored for specific density estimation problems, providing a better kernel approximation compared to a single-layer HEA architecture of equal or lesser depth.
However, the density estimation results using the DMKDE circuit reveal a limitation: the variational preparation of the training state relies on a fixed HEA architecture with a limited number of layers. This limitation hinders the circuit's scalability with respect to the number of qubits.
To overcome this, the vector kernel could be encoded directly without using kernel separability (as proposed in \cite{Woloshyn2024BoseHubbardMW}), allowing the maintenance of a fixed number of qubits while encoding multi-dimensional datasets. 
Overall, we obtain a DMKDE variational circuit that achieves an excellent approximation to 2D Gaussian KDE, when the number of qubits does not lead to expressivity issues in the HEA training circuit.

Future work for the implemented DMKDE circuit includes addressing scalability issues related to the number of qubits in the training circuit through vector data encoding and implementing various techniques to improve the circuit's trainability \cite{Cerezo2021CostFD, Grant2019AnIS, Shi2024AvoidingBP, Park2024HardwareefficientAW, Rad2022SurvivingTB, Volkoff2020LargeGV, Nakaji2020ExpressibilityOT}, conduct an implementation on real quantum hardware, study the expressiveness and structure of the ansätze found with genetic and memetic algorithms \cite{Nakaji2020ExpressibilityOT, Sim2019ExpressibilityAE}, and apply DMKDE circuits in other machine learning contexts where density estimation is used \cite{Bigdeli2020LearningGM, Nachman2020AnomalyDW, Hu2020AnomalyDU, Fraley2002ModelBasedCD, Bortoloti2020SupervisedKD}.

\section*{Acknowledgements}

J.E.A.-G. acknowledges funding from the project ``Aprendizaje de máquina para sistemas cuánticos'', HERMES code 57792, UNAL. H.V.-P. and F.A.G. acknowledge funding from the project ``Ampliación del uso de la mecánica cuántica desde el punto de vista experimental y su relación con la teoría, generando desarrollos en tecnologías cuánticas útiles para metrología y computación cuántica a nivel nacional'', BPIN 2022000100133, from SGR of MINCIENCIAS, Gobierno de Colombia. D.H.U. acknowledges Google Research for the support through the Google Ph.D. Fellowship.



\section*{Declarations}

\begin{itemize}
    \item \textbf{Competing Interests:} The authors declare the following competing interests: V.V.-C. was employed by Zapata Computing Inc. during the development of this work.
\end{itemize}

\section*{Data availability}

The implementation of MEMO-QCD and the datasets used in this work are available in the project repository: \url{https://gitlab.com/ml-physics-unal/memo-qcd}. 

\bibliography{main_bib}

\begin{thebibliography}{52}%
\makeatletter
\providecommand \@ifxundefined [1]{%
 \@ifx{#1\undefined}
}%
\providecommand \@ifnum [1]{%
 \ifnum #1\expandafter \@firstoftwo
 \else \expandafter \@secondoftwo
 \fi
}%
\providecommand \@ifx [1]{%
 \ifx #1\expandafter \@firstoftwo
 \else \expandafter \@secondoftwo
 \fi
}%
\providecommand \natexlab [1]{#1}%
\providecommand \enquote  [1]{``#1''}%
\providecommand \bibnamefont  [1]{#1}%
\providecommand \bibfnamefont [1]{#1}%
\providecommand \citenamefont [1]{#1}%
\providecommand \href@noop [0]{\@secondoftwo}%
\providecommand \href [0]{\begingroup \@sanitize@url \@href}%
\providecommand \@href[1]{\@@startlink{#1}\@@href}%
\providecommand \@@href[1]{\endgroup#1\@@endlink}%
\providecommand \@sanitize@url [0]{\catcode `\\12\catcode `\$12\catcode `\&12\catcode `\#12\catcode `\^12\catcode `\_12\catcode `\%12\relax}%
\providecommand \@@startlink[1]{}%
\providecommand \@@endlink[0]{}%
\providecommand \url  [0]{\begingroup\@sanitize@url \@url }%
\providecommand \@url [1]{\endgroup\@href {#1}{\urlprefix }}%
\providecommand \urlprefix  [0]{URL }%
\providecommand \Eprint [0]{\href }%
\providecommand \doibase [0]{https://doi.org/}%
\providecommand \selectlanguage [0]{\@gobble}%
\providecommand \bibinfo  [0]{\@secondoftwo}%
\providecommand \bibfield  [0]{\@secondoftwo}%
\providecommand \translation [1]{[#1]}%
\providecommand \BibitemOpen [0]{}%
\providecommand \bibitemStop [0]{}%
\providecommand \bibitemNoStop [0]{.\EOS\space}%
\providecommand \EOS [0]{\spacefactor3000\relax}%
\providecommand \BibitemShut  [1]{\csname bibitem#1\endcsname}%
\let\auto@bib@innerbib\@empty
\bibitem [{\citenamefont {Bigdeli}\ \emph {et~al.}(2020)\citenamefont {Bigdeli}, \citenamefont {Lin}, \citenamefont {Portenier}, \citenamefont {Dunbar},\ and\ \citenamefont {Zwicker}}]{Bigdeli2020LearningGM}%
  \BibitemOpen
  \bibfield  {author} {\bibinfo {author} {\bibfnamefont {S.~A.}\ \bibnamefont {Bigdeli}}, \bibinfo {author} {\bibfnamefont {G.}~\bibnamefont {Lin}}, \bibinfo {author} {\bibfnamefont {T.}~\bibnamefont {Portenier}}, \bibinfo {author} {\bibfnamefont {L.~A.}\ \bibnamefont {Dunbar}},\ and\ \bibinfo {author} {\bibfnamefont {M.}~\bibnamefont {Zwicker}},\ }\bibfield  {title} {\bibinfo {title} {Learning generative models using denoising density estimators},\ }\href {https://api.semanticscholar.org/CorpusID:210116656} {\bibfield  {journal} {\bibinfo  {journal} {IEEE transactions on neural networks and learning systems}\ }\textbf {\bibinfo {volume} {PP}} (\bibinfo {year} {2020})}\BibitemShut {NoStop}%
\bibitem [{\citenamefont {Nachman}\ and\ \citenamefont {Shih}(2020)}]{Nachman2020AnomalyDW}%
  \BibitemOpen
  \bibfield  {author} {\bibinfo {author} {\bibfnamefont {B.~P.}\ \bibnamefont {Nachman}}\ and\ \bibinfo {author} {\bibfnamefont {D.}~\bibnamefont {Shih}},\ }\bibfield  {title} {\bibinfo {title} {Anomaly detection with density estimation},\ }\href {https://api.semanticscholar.org/CorpusID:210718471} {\bibfield  {journal} {\bibinfo  {journal} {Physical Review D}\ } (\bibinfo {year} {2020})}\BibitemShut {NoStop}%
\bibitem [{\citenamefont {Hu}\ \emph {et~al.}(2020)\citenamefont {Hu}, \citenamefont {Gao}, \citenamefont {Li}, \citenamefont {Wu}, \citenamefont {Du},\ and\ \citenamefont {Maybank}}]{Hu2020AnomalyDU}%
  \BibitemOpen
  \bibfield  {author} {\bibinfo {author} {\bibfnamefont {W.}~\bibnamefont {Hu}}, \bibinfo {author} {\bibfnamefont {J.}~\bibnamefont {Gao}}, \bibinfo {author} {\bibfnamefont {B.}~\bibnamefont {Li}}, \bibinfo {author} {\bibfnamefont {O.}~\bibnamefont {Wu}}, \bibinfo {author} {\bibfnamefont {J.}~\bibnamefont {Du}},\ and\ \bibinfo {author} {\bibfnamefont {S.~J.}\ \bibnamefont {Maybank}},\ }\bibfield  {title} {\bibinfo {title} {Anomaly detection using local kernel density estimation and context-based regression},\ }\href {https://api.semanticscholar.org/CorpusID:69847830} {\bibfield  {journal} {\bibinfo  {journal} {IEEE Transactions on Knowledge and Data Engineering}\ }\textbf {\bibinfo {volume} {32}},\ \bibinfo {pages} {218} (\bibinfo {year} {2020})}\BibitemShut {NoStop}%
\bibitem [{\citenamefont {Fraley}\ and\ \citenamefont {Raftery}(2002)}]{Fraley2002ModelBasedCD}%
  \BibitemOpen
  \bibfield  {author} {\bibinfo {author} {\bibfnamefont {C.}~\bibnamefont {Fraley}}\ and\ \bibinfo {author} {\bibfnamefont {A.~E.}\ \bibnamefont {Raftery}},\ }\bibfield  {title} {\bibinfo {title} {Model-based clustering, discriminant analysis, and density estimation},\ }\href {https://api.semanticscholar.org/CorpusID:14462594} {\bibfield  {journal} {\bibinfo  {journal} {Journal of the American Statistical Association}\ }\textbf {\bibinfo {volume} {97}},\ \bibinfo {pages} {611 } (\bibinfo {year} {2002})}\BibitemShut {NoStop}%
\bibitem [{\citenamefont {Bortoloti}\ \emph {et~al.}(2020)\citenamefont {Bortoloti}, \citenamefont {de~Oliveira},\ and\ \citenamefont {Ciarelli}}]{Bortoloti2020SupervisedKD}%
  \BibitemOpen
  \bibfield  {author} {\bibinfo {author} {\bibfnamefont {F.~D.}\ \bibnamefont {Bortoloti}}, \bibinfo {author} {\bibfnamefont {E.}~\bibnamefont {de~Oliveira}},\ and\ \bibinfo {author} {\bibfnamefont {P.~M.}\ \bibnamefont {Ciarelli}},\ }\bibfield  {title} {\bibinfo {title} {Supervised kernel density estimation k-means},\ }\href {https://api.semanticscholar.org/CorpusID:229509441} {\bibfield  {journal} {\bibinfo  {journal} {Expert Syst. Appl.}\ }\textbf {\bibinfo {volume} {168}},\ \bibinfo {pages} {114350} (\bibinfo {year} {2020})}\BibitemShut {NoStop}%
\bibitem [{\citenamefont {Vapnik}\ and\ \citenamefont {Mukherjee}(1999)}]{Vapnik1999SupportVM}%
  \BibitemOpen
  \bibfield  {author} {\bibinfo {author} {\bibfnamefont {V.~N.}\ \bibnamefont {Vapnik}}\ and\ \bibinfo {author} {\bibfnamefont {S.}~\bibnamefont {Mukherjee}},\ }\bibfield  {title} {\bibinfo {title} {Support vector method for multivariate density estimation},\ }in\ \href {https://api.semanticscholar.org/CorpusID:1632365} {\emph {\bibinfo {booktitle} {Neural Information Processing Systems}}}\ (\bibinfo {year} {1999})\BibitemShut {NoStop}%
\bibitem [{\citenamefont {Papamakarios}\ \emph {et~al.}(2017)\citenamefont {Papamakarios}, \citenamefont {Murray},\ and\ \citenamefont {Pavlakou}}]{Papamakarios2017MaskedAF}%
  \BibitemOpen
  \bibfield  {author} {\bibinfo {author} {\bibfnamefont {G.}~\bibnamefont {Papamakarios}}, \bibinfo {author} {\bibfnamefont {I.}~\bibnamefont {Murray}},\ and\ \bibinfo {author} {\bibfnamefont {T.}~\bibnamefont {Pavlakou}},\ }\bibfield  {title} {\bibinfo {title} {Masked autoregressive flow for density estimation},\ }\href {https://api.semanticscholar.org/CorpusID:7166013} {\bibfield  {journal} {\bibinfo  {journal} {ArXiv}\ }\textbf {\bibinfo {volume} {abs/1705.07057}} (\bibinfo {year} {2017})}\BibitemShut {NoStop}%
\bibitem [{\citenamefont {Varanasi}\ and\ \citenamefont {Aazhang}(1989)}]{Varanasi1989ParametricGG}%
  \BibitemOpen
  \bibfield  {author} {\bibinfo {author} {\bibfnamefont {M.~K.}\ \bibnamefont {Varanasi}}\ and\ \bibinfo {author} {\bibfnamefont {B.}~\bibnamefont {Aazhang}},\ }\bibfield  {title} {\bibinfo {title} {Parametric generalized gaussian density estimation},\ }\href {https://api.semanticscholar.org/CorpusID:122172758} {\bibfield  {journal} {\bibinfo  {journal} {Journal of the Acoustical Society of America}\ }\textbf {\bibinfo {volume} {86}},\ \bibinfo {pages} {1404} (\bibinfo {year} {1989})}\BibitemShut {NoStop}%
\bibitem [{\citenamefont {Wang}\ and\ \citenamefont {Scott}(2019)}]{Wang2019NonparametricDE}%
  \BibitemOpen
  \bibfield  {author} {\bibinfo {author} {\bibfnamefont {Z.}~\bibnamefont {Wang}}\ and\ \bibinfo {author} {\bibfnamefont {D.~W.}\ \bibnamefont {Scott}},\ }\bibfield  {title} {\bibinfo {title} {Nonparametric density estimation for high‐dimensional data—algorithms and applications},\ }\href {https://api.semanticscholar.org/CorpusID:90262157} {\bibfield  {journal} {\bibinfo  {journal} {Wiley Interdisciplinary Reviews: Computational Statistics}\ }\textbf {\bibinfo {volume} {11}} (\bibinfo {year} {2019})}\BibitemShut {NoStop}%
\bibitem [{\citenamefont {Parzen}(1962)}]{Parzen1962OnEO}%
  \BibitemOpen
  \bibfield  {author} {\bibinfo {author} {\bibfnamefont {E.}~\bibnamefont {Parzen}},\ }\bibfield  {title} {\bibinfo {title} {On estimation of a probability density function and mode},\ }\href {https://api.semanticscholar.org/CorpusID:122932724} {\bibfield  {journal} {\bibinfo  {journal} {Annals of Mathematical Statistics}\ }\textbf {\bibinfo {volume} {33}},\ \bibinfo {pages} {1065} (\bibinfo {year} {1962})}\BibitemShut {NoStop}%
\bibitem [{\citenamefont {Rosenblatt}(1956)}]{Rosenblatt1956RemarksOS}%
  \BibitemOpen
  \bibfield  {author} {\bibinfo {author} {\bibfnamefont {M.}~\bibnamefont {Rosenblatt}},\ }\bibfield  {title} {\bibinfo {title} {Remarks on some nonparametric estimates of a density function},\ }\href {https://api.semanticscholar.org/CorpusID:16643156} {\bibfield  {journal} {\bibinfo  {journal} {Annals of Mathematical Statistics}\ }\textbf {\bibinfo {volume} {27}},\ \bibinfo {pages} {832} (\bibinfo {year} {1956})}\BibitemShut {NoStop}%
\bibitem [{\citenamefont {Liang}\ \emph {et~al.}(2019)\citenamefont {Liang}, \citenamefont {Shen}, \citenamefont {Li},\ and\ \citenamefont {Li}}]{Liang2019QuantumAD}%
  \BibitemOpen
  \bibfield  {author} {\bibinfo {author} {\bibfnamefont {J.}~\bibnamefont {Liang}}, \bibinfo {author} {\bibfnamefont {S.}~\bibnamefont {Shen}}, \bibinfo {author} {\bibfnamefont {M.}~\bibnamefont {Li}},\ and\ \bibinfo {author} {\bibfnamefont {L.}~\bibnamefont {Li}},\ }\bibfield  {title} {\bibinfo {title} {Quantum anomaly detection with density estimation and multivariate gaussian distribution},\ }\href {https://api.semanticscholar.org/CorpusID:164885166} {\bibfield  {journal} {\bibinfo  {journal} {Physical Review A}\ } (\bibinfo {year} {2019})}\BibitemShut {NoStop}%
\bibitem [{\citenamefont {Guo}\ \emph {et~al.}(2021)\citenamefont {Guo}, \citenamefont {Liu}, \citenamefont {Li}, \citenamefont {Li}, \citenamefont {Qin}, \citenamefont {Wen},\ and\ \citenamefont {Gao}}]{Guo2021QuantumAF}%
  \BibitemOpen
  \bibfield  {author} {\bibinfo {author} {\bibfnamefont {M.}~\bibnamefont {Guo}}, \bibinfo {author} {\bibfnamefont {H.}~\bibnamefont {Liu}}, \bibinfo {author} {\bibfnamefont {Y.}~\bibnamefont {Li}}, \bibinfo {author} {\bibfnamefont {W.}~\bibnamefont {Li}}, \bibinfo {author} {\bibfnamefont {S.}~\bibnamefont {Qin}}, \bibinfo {author} {\bibfnamefont {Q.}~\bibnamefont {Wen}},\ and\ \bibinfo {author} {\bibfnamefont {F.}~\bibnamefont {Gao}},\ }\bibfield  {title} {\bibinfo {title} {Quantum algorithms for anomaly detection using amplitude estimation},\ }\href {https://api.semanticscholar.org/CorpusID:238198477} {\bibfield  {journal} {\bibinfo  {journal} {SSRN Electronic Journal}\ } (\bibinfo {year} {2021})}\BibitemShut {NoStop}%
\bibitem [{\citenamefont {Verdon}\ \emph {et~al.}(2019)\citenamefont {Verdon}, \citenamefont {Marks}, \citenamefont {Nanda}, \citenamefont {Leichenauer},\ and\ \citenamefont {Hidary}}]{Verdon2019QuantumHM}%
  \BibitemOpen
  \bibfield  {author} {\bibinfo {author} {\bibfnamefont {G.}~\bibnamefont {Verdon}}, \bibinfo {author} {\bibfnamefont {J.~A.}\ \bibnamefont {Marks}}, \bibinfo {author} {\bibfnamefont {S.}~\bibnamefont {Nanda}}, \bibinfo {author} {\bibfnamefont {S.}~\bibnamefont {Leichenauer}},\ and\ \bibinfo {author} {\bibfnamefont {J.~D.}\ \bibnamefont {Hidary}},\ }\bibfield  {title} {\bibinfo {title} {Quantum hamiltonian-based models and the variational quantum thermalizer algorithm},\ }\href {https://api.semanticscholar.org/CorpusID:203736873} {\bibfield  {journal} {\bibinfo  {journal} {ArXiv}\ }\textbf {\bibinfo {volume} {abs/1910.02071}} (\bibinfo {year} {2019})}\BibitemShut {NoStop}%
\bibitem [{\citenamefont {Araz}\ and\ \citenamefont {Spannowsky}(2023)}]{PhysRevA.108.062422}%
  \BibitemOpen
  \bibfield  {author} {\bibinfo {author} {\bibfnamefont {J.~Y.}\ \bibnamefont {Araz}}\ and\ \bibinfo {author} {\bibfnamefont {M.}~\bibnamefont {Spannowsky}},\ }\bibfield  {title} {\bibinfo {title} {Quantum-probabilistic hamiltonian learning for generative modeling and anomaly detection},\ }\href {https://doi.org/10.1103/PhysRevA.108.062422} {\bibfield  {journal} {\bibinfo  {journal} {Phys. Rev. A}\ }\textbf {\bibinfo {volume} {108}},\ \bibinfo {pages} {062422} (\bibinfo {year} {2023})}\BibitemShut {NoStop}%
\bibitem [{\citenamefont {Schuhmacher}\ \emph {et~al.}(2023)\citenamefont {Schuhmacher}, \citenamefont {Boggia}, \citenamefont {Belis}, \citenamefont {Puljak}, \citenamefont {Grossi}, \citenamefont {Pierini}, \citenamefont {Vallecorsa}, \citenamefont {Tacchino}, \citenamefont {Barkoutsos},\ and\ \citenamefont {Tavernelli}}]{Schuhmacher2023UnravellingPB}%
  \BibitemOpen
  \bibfield  {author} {\bibinfo {author} {\bibfnamefont {J.}~\bibnamefont {Schuhmacher}}, \bibinfo {author} {\bibfnamefont {L.}~\bibnamefont {Boggia}}, \bibinfo {author} {\bibfnamefont {V.}~\bibnamefont {Belis}}, \bibinfo {author} {\bibfnamefont {E.}~\bibnamefont {Puljak}}, \bibinfo {author} {\bibfnamefont {M.}~\bibnamefont {Grossi}}, \bibinfo {author} {\bibfnamefont {M.}~\bibnamefont {Pierini}}, \bibinfo {author} {\bibfnamefont {S.}~\bibnamefont {Vallecorsa}}, \bibinfo {author} {\bibfnamefont {F.}~\bibnamefont {Tacchino}}, \bibinfo {author} {\bibfnamefont {P.~K.}\ \bibnamefont {Barkoutsos}},\ and\ \bibinfo {author} {\bibfnamefont {I.}~\bibnamefont {Tavernelli}},\ }\bibfield  {title} {\bibinfo {title} {Unravelling physics beyond the standard model with classical and quantum anomaly detection},\ }\href {https://api.semanticscholar.org/CorpusID:256274836} {\bibfield  {journal} {\bibinfo  {journal} {Machine Learning: Science and Technology}\ }\textbf {\bibinfo {volume} {4}} (\bibinfo {year} {2023})}\BibitemShut
  {NoStop}%
\bibitem [{\citenamefont {González}\ \emph {et~al.}(2020)\citenamefont {González}, \citenamefont {Vargas-Calderón},\ and\ \citenamefont {Vinck-Posada}}]{Gonzlez2020ClassificationWQ}%
  \BibitemOpen
  \bibfield  {author} {\bibinfo {author} {\bibfnamefont {F.~A.}\ \bibnamefont {González}}, \bibinfo {author} {\bibfnamefont {V.}~\bibnamefont {Vargas-Calderón}},\ and\ \bibinfo {author} {\bibfnamefont {H.}~\bibnamefont {Vinck-Posada}},\ }\bibfield  {title} {\bibinfo {title} {Classification with quantum measurements},\ }\href {https://api.semanticscholar.org/CorpusID:233697843} {\bibfield  {journal} {\bibinfo  {journal} {Journal of the Physical Society of Japan}\ }\textbf {\bibinfo {volume} {90}},\ \bibinfo {pages} {044002} (\bibinfo {year} {2020})}\BibitemShut {NoStop}%
\bibitem [{\citenamefont {Useche}\ \emph {et~al.}(2021)\citenamefont {Useche}, \citenamefont {Giraldo-Carvajal}, \citenamefont {Zuluaga-Bucheli}, \citenamefont {Jaramillo-Villegas},\ and\ \citenamefont {González}}]{Useche2021QuantumMC}%
  \BibitemOpen
  \bibfield  {author} {\bibinfo {author} {\bibfnamefont {D.~H.}\ \bibnamefont {Useche}}, \bibinfo {author} {\bibfnamefont {A.}~\bibnamefont {Giraldo-Carvajal}}, \bibinfo {author} {\bibfnamefont {H.~M.}\ \bibnamefont {Zuluaga-Bucheli}}, \bibinfo {author} {\bibfnamefont {J.~A.}\ \bibnamefont {Jaramillo-Villegas}},\ and\ \bibinfo {author} {\bibfnamefont {F.~A.}\ \bibnamefont {González}},\ }\bibfield  {title} {\bibinfo {title} {Quantum measurement classification with qudits},\ }\href {https://api.semanticscholar.org/CorpusID:236154740} {\bibfield  {journal} {\bibinfo  {journal} {Quantum Information Processing}\ }\textbf {\bibinfo {volume} {21}} (\bibinfo {year} {2021})}\BibitemShut {NoStop}%
\bibitem [{\citenamefont {González}\ \emph {et~al.}(2021)\citenamefont {González}, \citenamefont {Gallego-Mejia}, \citenamefont {Toledo-Cortés},\ and\ \citenamefont {Vargas-Calderón}}]{Gonzalez2021LearningWD}%
  \BibitemOpen
  \bibfield  {author} {\bibinfo {author} {\bibfnamefont {F.~A.}\ \bibnamefont {González}}, \bibinfo {author} {\bibfnamefont {J.~A.}\ \bibnamefont {Gallego-Mejia}}, \bibinfo {author} {\bibfnamefont {S.}~\bibnamefont {Toledo-Cortés}},\ and\ \bibinfo {author} {\bibfnamefont {V.}~\bibnamefont {Vargas-Calderón}},\ }\bibfield  {title} {\bibinfo {title} {Learning with density matrices and random features},\ }\href {https://api.semanticscholar.org/CorpusID:231847224} {\bibfield  {journal} {\bibinfo  {journal} {Quantum Machine Intelligence}\ }\textbf {\bibinfo {volume} {4}} (\bibinfo {year} {2021})}\BibitemShut {NoStop}%
\bibitem [{\citenamefont {Nakayama}\ \emph {et~al.}(2024)\citenamefont {Nakayama}, \citenamefont {Morisaki}, \citenamefont {Mitarai}, \citenamefont {Ueda},\ and\ \citenamefont {Fujii}}]{nakayama2024explicitquantumsurrogatesquantum}%
  \BibitemOpen
  \bibfield  {author} {\bibinfo {author} {\bibfnamefont {A.}~\bibnamefont {Nakayama}}, \bibinfo {author} {\bibfnamefont {H.}~\bibnamefont {Morisaki}}, \bibinfo {author} {\bibfnamefont {K.}~\bibnamefont {Mitarai}}, \bibinfo {author} {\bibfnamefont {H.}~\bibnamefont {Ueda}},\ and\ \bibinfo {author} {\bibfnamefont {K.}~\bibnamefont {Fujii}},\ }\href {https://arxiv.org/abs/2408.03000} {\bibinfo {title} {Explicit quantum surrogates for quantum kernel models}} (\bibinfo {year} {2024}),\ \Eprint {https://arxiv.org/abs/2408.03000} {arXiv:2408.03000 [quant-ph]} \BibitemShut {NoStop}%
\bibitem [{\citenamefont {Gallego-Mejia}\ \emph {et~al.}(2022{\natexlab{a}})\citenamefont {Gallego-Mejia}, \citenamefont {Bustos-Brinez},\ and\ \citenamefont {González}}]{GallegoMejia2022InQMADIQ}%
  \BibitemOpen
  \bibfield  {author} {\bibinfo {author} {\bibfnamefont {J.~A.}\ \bibnamefont {Gallego-Mejia}}, \bibinfo {author} {\bibfnamefont {O.~A.}\ \bibnamefont {Bustos-Brinez}},\ and\ \bibinfo {author} {\bibfnamefont {F.}~\bibnamefont {González}},\ }\bibfield  {title} {\bibinfo {title} {Inqmad: Incremental quantum measurement anomaly detection},\ }\href {https://api.semanticscholar.org/CorpusID:252815662} {\bibfield  {journal} {\bibinfo  {journal} {2022 IEEE International Conference on Data Mining Workshops (ICDMW)}\ ,\ \bibinfo {pages} {787}} (\bibinfo {year} {2022}{\natexlab{a}})}\BibitemShut {NoStop}%
\bibitem [{\citenamefont {Gallego-Mejia}\ \emph {et~al.}(2022{\natexlab{b}})\citenamefont {Gallego-Mejia}, \citenamefont {Osorio},\ and\ \citenamefont {González}}]{GallegoMejia2022FastKD}%
  \BibitemOpen
  \bibfield  {author} {\bibinfo {author} {\bibfnamefont {J.~A.}\ \bibnamefont {Gallego-Mejia}}, \bibinfo {author} {\bibfnamefont {J.~F.}\ \bibnamefont {Osorio}},\ and\ \bibinfo {author} {\bibfnamefont {F.~A.}\ \bibnamefont {González}},\ }\bibfield  {title} {\bibinfo {title} {Fast kernel density estimation with density matrices and random fourier features},\ }\href {https://api.semanticscholar.org/CorpusID:251253329} {\bibfield  {journal} {\bibinfo  {journal} {ArXiv}\ }\textbf {\bibinfo {volume} {abs/2208.01206}} (\bibinfo {year} {2022}{\natexlab{b}})}\BibitemShut {NoStop}%
\bibitem [{\citenamefont {Gallego-Mejia}\ and\ \citenamefont {González}(2022)}]{GallegoMejia2022QuantumAF}%
  \BibitemOpen
  \bibfield  {author} {\bibinfo {author} {\bibfnamefont {J.~A.}\ \bibnamefont {Gallego-Mejia}}\ and\ \bibinfo {author} {\bibfnamefont {F.~A.}\ \bibnamefont {González}},\ }\bibfield  {title} {\bibinfo {title} {Quantum adaptive fourier features for neural density estimation},\ }\href {https://api.semanticscholar.org/CorpusID:251224212} {\bibfield  {journal} {\bibinfo  {journal} {ArXiv}\ }\textbf {\bibinfo {volume} {abs/2208.00564}} (\bibinfo {year} {2022})}\BibitemShut {NoStop}%
\bibitem [{\citenamefont {Gallego-Mejia}\ \emph {et~al.}(2022{\natexlab{c}})\citenamefont {Gallego-Mejia}, \citenamefont {Bustos-Brinez},\ and\ \citenamefont {González}}]{GallegoMejia2022LEANDMKDEQL}%
  \BibitemOpen
  \bibfield  {author} {\bibinfo {author} {\bibfnamefont {J.~A.}\ \bibnamefont {Gallego-Mejia}}, \bibinfo {author} {\bibfnamefont {O.~A.}\ \bibnamefont {Bustos-Brinez}},\ and\ \bibinfo {author} {\bibfnamefont {F.~A.}\ \bibnamefont {González}},\ }\bibfield  {title} {\bibinfo {title} {Lean-dmkde: Quantum latent density estimation for anomaly detection},\ }in\ \href {https://api.semanticscholar.org/CorpusID:253553422} {\emph {\bibinfo {booktitle} {AAAI Conference on Artificial Intelligence}}}\ (\bibinfo {year} {2022})\BibitemShut {NoStop}%
\bibitem [{\citenamefont {Bustos-Brinez}\ \emph {et~al.}(2022)\citenamefont {Bustos-Brinez}, \citenamefont {Gallego-Mejia},\ and\ \citenamefont {González}}]{BustosBrinez2022ADDMKDEAD}%
  \BibitemOpen
  \bibfield  {author} {\bibinfo {author} {\bibfnamefont {O.~A.}\ \bibnamefont {Bustos-Brinez}}, \bibinfo {author} {\bibfnamefont {J.~A.}\ \bibnamefont {Gallego-Mejia}},\ and\ \bibinfo {author} {\bibfnamefont {F.~A.}\ \bibnamefont {González}},\ }\bibfield  {title} {\bibinfo {title} {Ad-dmkde: Anomaly detection through density matrices and fourier features},\ }\href {https://api.semanticscholar.org/CorpusID:253116656} {\bibfield  {journal} {\bibinfo  {journal} {ArXiv}\ }\textbf {\bibinfo {volume} {abs/2210.14796}} (\bibinfo {year} {2022})}\BibitemShut {NoStop}%
\bibitem [{\citenamefont {Toledo-Cortés}\ \emph {et~al.}(2022)\citenamefont {Toledo-Cortés}, \citenamefont {Useche}, \citenamefont {M{\"u}ller},\ and\ \citenamefont {González}}]{ToledoCorts2022GradingDR}%
  \BibitemOpen
  \bibfield  {author} {\bibinfo {author} {\bibfnamefont {S.}~\bibnamefont {Toledo-Cortés}}, \bibinfo {author} {\bibfnamefont {D.~H.}\ \bibnamefont {Useche}}, \bibinfo {author} {\bibfnamefont {H.}~\bibnamefont {M{\"u}ller}},\ and\ \bibinfo {author} {\bibfnamefont {F.~A.}\ \bibnamefont {González}},\ }\bibfield  {title} {\bibinfo {title} {Grading diabetic retinopathy and prostate cancer diagnostic images with deep quantum ordinal regression},\ }\href {https://api.semanticscholar.org/CorpusID:248075461} {\bibfield  {journal} {\bibinfo  {journal} {Computers in biology and medicine}\ }\textbf {\bibinfo {volume} {145}},\ \bibinfo {pages} {105472} (\bibinfo {year} {2022})}\BibitemShut {NoStop}%
\bibitem [{\citenamefont {Vargas-Calderón}\ \emph {et~al.}(2022)\citenamefont {Vargas-Calderón}, \citenamefont {González},\ and\ \citenamefont {Vinck-Posada}}]{VargasCalderon2022OptimisationfreeDE}%
  \BibitemOpen
  \bibfield  {author} {\bibinfo {author} {\bibfnamefont {V.}~\bibnamefont {Vargas-Calderón}}, \bibinfo {author} {\bibfnamefont {F.~A.}\ \bibnamefont {González}},\ and\ \bibinfo {author} {\bibfnamefont {H.}~\bibnamefont {Vinck-Posada}},\ }\bibfield  {title} {\bibinfo {title} {Optimisation-free density estimation and classification with quantum circuits},\ }\href {https://api.semanticscholar.org/CorpusID:247762127} {\bibfield  {journal} {\bibinfo  {journal} {Quantum Machine Intelligence}\ }\textbf {\bibinfo {volume} {4}} (\bibinfo {year} {2022})}\BibitemShut {NoStop}%
\bibitem [{\citenamefont {Moscato}(1999)}]{Moscato1999MemeticAA}%
  \BibitemOpen
  \bibfield  {author} {\bibinfo {author} {\bibfnamefont {P.}~\bibnamefont {Moscato}},\ }\bibfield  {title} {\bibinfo {title} {Memetic algorithms: a short introduction}\ }(\bibinfo {year} {1999})\BibitemShut {NoStop}%
\bibitem [{\citenamefont {Altares-López}\ \emph {et~al.}(2021)\citenamefont {Altares-López}, \citenamefont {Ribeiro},\ and\ \citenamefont {García-Ripoll}}]{AltaresLpez2021AutomaticDO}%
  \BibitemOpen
  \bibfield  {author} {\bibinfo {author} {\bibfnamefont {S.}~\bibnamefont {Altares-López}}, \bibinfo {author} {\bibfnamefont {{\'A}.}~\bibnamefont {Ribeiro}},\ and\ \bibinfo {author} {\bibfnamefont {J.~J.}\ \bibnamefont {García-Ripoll}},\ }\bibfield  {title} {\bibinfo {title} {Automatic design of quantum feature maps},\ }\href {https://api.semanticscholar.org/CorpusID:235195771} {\bibfield  {journal} {\bibinfo  {journal} {Quantum Science \& Technology}\ }\textbf {\bibinfo {volume} {6}} (\bibinfo {year} {2021})}\BibitemShut {NoStop}%
\bibitem [{\citenamefont {Ruder}(2016)}]{Ruder2016AnOO}%
  \BibitemOpen
  \bibfield  {author} {\bibinfo {author} {\bibfnamefont {S.}~\bibnamefont {Ruder}},\ }\bibfield  {title} {\bibinfo {title} {An overview of gradient descent optimization algorithms},\ }\href {https://api.semanticscholar.org/CorpusID:17485266} {\bibfield  {journal} {\bibinfo  {journal} {ArXiv}\ }\textbf {\bibinfo {volume} {abs/1609.04747}} (\bibinfo {year} {2016})}\BibitemShut {NoStop}%
\bibitem [{\citenamefont {Cerezo}\ \emph {et~al.}(2020)\citenamefont {Cerezo}, \citenamefont {Arrasmith}, \citenamefont {Babbush}, \citenamefont {Benjamin}, \citenamefont {Endo}, \citenamefont {Fujii}, \citenamefont {McClean}, \citenamefont {Mitarai}, \citenamefont {Yuan}, \citenamefont {Cincio},\ and\ \citenamefont {Coles}}]{Cerezo2020VariationalQA}%
  \BibitemOpen
  \bibfield  {author} {\bibinfo {author} {\bibfnamefont {M.}~\bibnamefont {Cerezo}}, \bibinfo {author} {\bibfnamefont {A.}~\bibnamefont {Arrasmith}}, \bibinfo {author} {\bibfnamefont {R.}~\bibnamefont {Babbush}}, \bibinfo {author} {\bibfnamefont {S.~C.}\ \bibnamefont {Benjamin}}, \bibinfo {author} {\bibfnamefont {S.}~\bibnamefont {Endo}}, \bibinfo {author} {\bibfnamefont {K.}~\bibnamefont {Fujii}}, \bibinfo {author} {\bibfnamefont {J.~R.}\ \bibnamefont {McClean}}, \bibinfo {author} {\bibfnamefont {K.}~\bibnamefont {Mitarai}}, \bibinfo {author} {\bibfnamefont {X.}~\bibnamefont {Yuan}}, \bibinfo {author} {\bibfnamefont {L.}~\bibnamefont {Cincio}},\ and\ \bibinfo {author} {\bibfnamefont {P.~J.}\ \bibnamefont {Coles}},\ }\bibfield  {title} {\bibinfo {title} {Variational quantum algorithms},\ }\href {https://api.semanticscholar.org/CorpusID:229297850} {\bibfield  {journal} {\bibinfo  {journal} {Nature Reviews Physics}\ }\textbf {\bibinfo {volume} {3}},\ \bibinfo {pages} {625 } (\bibinfo {year} {2020})}\BibitemShut
  {NoStop}%
\bibitem [{\citenamefont {Kandala}\ \emph {et~al.}(2017)\citenamefont {Kandala}, \citenamefont {Mezzacapo}, \citenamefont {Temme}, \citenamefont {Takita}, \citenamefont {Brink}, \citenamefont {Chow},\ and\ \citenamefont {Gambetta}}]{Kandala2017HardwareefficientVQ}%
  \BibitemOpen
  \bibfield  {author} {\bibinfo {author} {\bibfnamefont {A.}~\bibnamefont {Kandala}}, \bibinfo {author} {\bibfnamefont {A.}~\bibnamefont {Mezzacapo}}, \bibinfo {author} {\bibfnamefont {K.}~\bibnamefont {Temme}}, \bibinfo {author} {\bibfnamefont {M.}~\bibnamefont {Takita}}, \bibinfo {author} {\bibfnamefont {M.}~\bibnamefont {Brink}}, \bibinfo {author} {\bibfnamefont {J.~M.}\ \bibnamefont {Chow}},\ and\ \bibinfo {author} {\bibfnamefont {J.~M.}\ \bibnamefont {Gambetta}},\ }\bibfield  {title} {\bibinfo {title} {Hardware-efficient variational quantum eigensolver for small molecules and quantum magnets},\ }\href {https://api.semanticscholar.org/CorpusID:4390182} {\bibfield  {journal} {\bibinfo  {journal} {Nature}\ }\textbf {\bibinfo {volume} {549}},\ \bibinfo {pages} {242} (\bibinfo {year} {2017})}\BibitemShut {NoStop}%
\bibitem [{\citenamefont {Preskill}(2018)}]{Preskill2018QuantumCI}%
  \BibitemOpen
  \bibfield  {author} {\bibinfo {author} {\bibfnamefont {J.}~\bibnamefont {Preskill}},\ }\bibfield  {title} {\bibinfo {title} {Quantum computing in the nisq era and beyond},\ }\href {https://api.semanticscholar.org/CorpusID:44098998} {\bibfield  {journal} {\bibinfo  {journal} {Quantum}\ } (\bibinfo {year} {2018})}\BibitemShut {NoStop}%
\bibitem [{\citenamefont {Brandhofer}\ \emph {et~al.}(2021)\citenamefont {Brandhofer}, \citenamefont {Devitt}, \citenamefont {Wellens},\ and\ \citenamefont {Polian}}]{Brandhofer2021SpecialSN}%
  \BibitemOpen
  \bibfield  {author} {\bibinfo {author} {\bibfnamefont {S.}~\bibnamefont {Brandhofer}}, \bibinfo {author} {\bibfnamefont {S.~J.}\ \bibnamefont {Devitt}}, \bibinfo {author} {\bibfnamefont {T.}~\bibnamefont {Wellens}},\ and\ \bibinfo {author} {\bibfnamefont {I.}~\bibnamefont {Polian}},\ }\bibfield  {title} {\bibinfo {title} {Special session: Noisy intermediate-scale quantum (nisq) computers—how they work, how they fail, how to test them?},\ }\href {https://api.semanticscholar.org/CorpusID:235307511} {\bibfield  {journal} {\bibinfo  {journal} {2021 IEEE 39th VLSI Test Symposium (VTS)}\ ,\ \bibinfo {pages} {1}} (\bibinfo {year} {2021})}\BibitemShut {NoStop}%
\bibitem [{\citenamefont {Useche}\ \emph {et~al.}(2022)\citenamefont {Useche}, \citenamefont {Bustos-Brinez}, \citenamefont {Gallego-Mejia},\ and\ \citenamefont {González}}]{Useche2022QuantumDE}%
  \BibitemOpen
  \bibfield  {author} {\bibinfo {author} {\bibfnamefont {D.~H.}\ \bibnamefont {Useche}}, \bibinfo {author} {\bibfnamefont {O.~A.}\ \bibnamefont {Bustos-Brinez}}, \bibinfo {author} {\bibfnamefont {J.~A.}\ \bibnamefont {Gallego-Mejia}},\ and\ \bibinfo {author} {\bibfnamefont {F.~A.}\ \bibnamefont {González}},\ }\bibfield  {title} {\bibinfo {title} {Quantum density estimation with density matrices: Application to quantum anomaly detection},\ }\href {https://api.semanticscholar.org/CorpusID:266521602} {\bibfield  {journal} {\bibinfo  {journal} {Physical Review A}\ } (\bibinfo {year} {2022})}\BibitemShut {NoStop}%
\bibitem [{Note1()}]{Note1}%
  \BibitemOpen
  \bibinfo {note} {This is called a training state because building it resembles the usual training step in machine learning methods. However, in principle, there does not need to be an actual training procedure to build such training state.}\BibitemShut {Stop}%
\bibitem [{\citenamefont {Katoch}\ \emph {et~al.}(2020)\citenamefont {Katoch}, \citenamefont {Chauhan},\ and\ \citenamefont {Kumar}}]{Katoch2020ARO}%
  \BibitemOpen
  \bibfield  {author} {\bibinfo {author} {\bibfnamefont {S.}~\bibnamefont {Katoch}}, \bibinfo {author} {\bibfnamefont {S.~S.}\ \bibnamefont {Chauhan}},\ and\ \bibinfo {author} {\bibfnamefont {V.}~\bibnamefont {Kumar}},\ }\bibfield  {title} {\bibinfo {title} {A review on genetic algorithm: past, present, and future},\ }\href {https://api.semanticscholar.org/CorpusID:226227415} {\bibfield  {journal} {\bibinfo  {journal} {Multimedia Tools and Applications}\ }\textbf {\bibinfo {volume} {80}},\ \bibinfo {pages} {8091 } (\bibinfo {year} {2020})}\BibitemShut {NoStop}%
\bibitem [{Note2()}]{Note2}%
  \BibitemOpen
  \bibinfo {note} {As in the case of finding the QFM circuit, to find the training state circuit we also used genetic and memetic algorithms for optimising the quantum circuit architecture, but got little to no improvement with respect to the shallow HEA architecture (which showed good results) while incurring in a significant computational time overhead.}\BibitemShut {Stop}%
\bibitem [{Note3()}]{Note3}%
  \BibitemOpen
  \bibinfo {note} {Due to its superior performance, only the QFM circuit found by the memetic algorithm is used to build the training circuit, and consequently, the DMKDE circuit.}\BibitemShut {Stop}%
\bibitem [{\citenamefont {P{\'e}rez-Cruz}(2008)}]{PrezCruz2008KullbackLeiblerDE}%
  \BibitemOpen
  \bibfield  {author} {\bibinfo {author} {\bibfnamefont {F.}~\bibnamefont {P{\'e}rez-Cruz}},\ }\bibfield  {title} {\bibinfo {title} {Kullback-leibler divergence estimation of continuous distributions},\ }\href {https://api.semanticscholar.org/CorpusID:8811865} {\bibfield  {journal} {\bibinfo  {journal} {2008 IEEE International Symposium on Information Theory}\ ,\ \bibinfo {pages} {1666}} (\bibinfo {year} {2008})}\BibitemShut {NoStop}%
\bibitem [{\citenamefont {Bowles}\ \emph {et~al.}(2024)\citenamefont {Bowles}, \citenamefont {Ahmed},\ and\ \citenamefont {Schuld}}]{bowles2024better}%
  \BibitemOpen
  \bibfield  {author} {\bibinfo {author} {\bibfnamefont {J.}~\bibnamefont {Bowles}}, \bibinfo {author} {\bibfnamefont {S.}~\bibnamefont {Ahmed}},\ and\ \bibinfo {author} {\bibfnamefont {M.}~\bibnamefont {Schuld}},\ }\href@noop {} {\bibinfo {title} {Better than classical? the subtle art of benchmarking quantum machine learning models}} (\bibinfo {year} {2024}),\ \Eprint {https://arxiv.org/abs/2403.07059} {arXiv:2403.07059 [quant-ph]} \BibitemShut {NoStop}%
\bibitem [{\citenamefont {Nakaji}\ and\ \citenamefont {Yamamoto}(2020)}]{Nakaji2020ExpressibilityOT}%
  \BibitemOpen
  \bibfield  {author} {\bibinfo {author} {\bibfnamefont {K.}~\bibnamefont {Nakaji}}\ and\ \bibinfo {author} {\bibfnamefont {N.}~\bibnamefont {Yamamoto}},\ }\bibfield  {title} {\bibinfo {title} {Expressibility of the alternating layered ansatz for quantum computation},\ }\href {https://api.semanticscholar.org/CorpusID:218889362} {\bibfield  {journal} {\bibinfo  {journal} {Quantum}\ }\textbf {\bibinfo {volume} {5}},\ \bibinfo {pages} {434} (\bibinfo {year} {2020})}\BibitemShut {NoStop}%
\bibitem [{\citenamefont {Sim}\ \emph {et~al.}(2019)\citenamefont {Sim}, \citenamefont {Johnson},\ and\ \citenamefont {Aspuru-Guzik}}]{Sim2019ExpressibilityAE}%
  \BibitemOpen
  \bibfield  {author} {\bibinfo {author} {\bibfnamefont {S.}~\bibnamefont {Sim}}, \bibinfo {author} {\bibfnamefont {P.~D.}\ \bibnamefont {Johnson}},\ and\ \bibinfo {author} {\bibfnamefont {A.}~\bibnamefont {Aspuru-Guzik}},\ }\bibfield  {title} {\bibinfo {title} {Expressibility and entangling capability of parameterized quantum circuits for hybrid quantum‐classical algorithms},\ }\href {https://api.semanticscholar.org/CorpusID:166228228} {\bibfield  {journal} {\bibinfo  {journal} {Advanced Quantum Technologies}\ }\textbf {\bibinfo {volume} {2}} (\bibinfo {year} {2019})}\BibitemShut {NoStop}%
\bibitem [{\citenamefont {Woloshyn}\ and\ \citenamefont {Mall}(2024)}]{Woloshyn2024BoseHubbardMW}%
  \BibitemOpen
  \bibfield  {author} {\bibinfo {author} {\bibfnamefont {R.~M.}\ \bibnamefont {Woloshyn}}\ and\ \bibinfo {author} {\bibfnamefont {W.}~\bibnamefont {Mall}},\ }\bibfield  {title} {\bibinfo {title} {Bose-hubbard model with a single qubit}\ }(\bibinfo {year} {2024})\BibitemShut {NoStop}%
\bibitem [{\citenamefont {Cerezo}\ \emph {et~al.}(2021)\citenamefont {Cerezo}, \citenamefont {Sone}, \citenamefont {Volkoff}, \citenamefont {Cincio},\ and\ \citenamefont {Coles}}]{Cerezo2021CostFD}%
  \BibitemOpen
  \bibfield  {author} {\bibinfo {author} {\bibfnamefont {M.}~\bibnamefont {Cerezo}}, \bibinfo {author} {\bibfnamefont {A.}~\bibnamefont {Sone}}, \bibinfo {author} {\bibfnamefont {T.~J.}\ \bibnamefont {Volkoff}}, \bibinfo {author} {\bibfnamefont {L.}~\bibnamefont {Cincio}},\ and\ \bibinfo {author} {\bibfnamefont {P.~J.}\ \bibnamefont {Coles}},\ }\bibfield  {title} {\bibinfo {title} {Cost function dependent barren plateaus in shallow parametrized quantum circuits},\ }\href {https://api.semanticscholar.org/CorpusID:232298392} {\bibfield  {journal} {\bibinfo  {journal} {Nature Communications}\ }\textbf {\bibinfo {volume} {12}} (\bibinfo {year} {2021})}\BibitemShut {NoStop}%
\bibitem [{\citenamefont {Grant}\ \emph {et~al.}(2019)\citenamefont {Grant}, \citenamefont {Wossnig}, \citenamefont {Ostaszewski},\ and\ \citenamefont {Benedetti}}]{Grant2019AnIS}%
  \BibitemOpen
  \bibfield  {author} {\bibinfo {author} {\bibfnamefont {E.}~\bibnamefont {Grant}}, \bibinfo {author} {\bibfnamefont {L.}~\bibnamefont {Wossnig}}, \bibinfo {author} {\bibfnamefont {M.}~\bibnamefont {Ostaszewski}},\ and\ \bibinfo {author} {\bibfnamefont {M.}~\bibnamefont {Benedetti}},\ }\bibfield  {title} {\bibinfo {title} {An initialization strategy for addressing barren plateaus in parametrized quantum circuits},\ }\href {https://api.semanticscholar.org/CorpusID:92984249} {\bibfield  {journal} {\bibinfo  {journal} {Quantum}\ } (\bibinfo {year} {2019})}\BibitemShut {NoStop}%
\bibitem [{\citenamefont {Shi}\ and\ \citenamefont {Shang}(2024)}]{Shi2024AvoidingBP}%
  \BibitemOpen
  \bibfield  {author} {\bibinfo {author} {\bibfnamefont {X.}~\bibnamefont {Shi}}\ and\ \bibinfo {author} {\bibfnamefont {Y.}~\bibnamefont {Shang}},\ }\bibfield  {title} {\bibinfo {title} {Avoiding barren plateaus via gaussian mixture model}\ }(\bibinfo {year} {2024})\BibitemShut {NoStop}%
\bibitem [{\citenamefont {Park}\ \emph {et~al.}(2024)\citenamefont {Park}, \citenamefont {Kang},\ and\ \citenamefont {Huh}}]{Park2024HardwareefficientAW}%
  \BibitemOpen
  \bibfield  {author} {\bibinfo {author} {\bibfnamefont {C.-Y.}\ \bibnamefont {Park}}, \bibinfo {author} {\bibfnamefont {M.}~\bibnamefont {Kang}},\ and\ \bibinfo {author} {\bibfnamefont {J.}~\bibnamefont {Huh}},\ }\bibfield  {title} {\bibinfo {title} {Hardware-efficient ansatz without barren plateaus in any depth}\ }(\bibinfo {year} {2024})\BibitemShut {NoStop}%
\bibitem [{\citenamefont {Rad}\ \emph {et~al.}(2022)\citenamefont {Rad}, \citenamefont {Seif},\ and\ \citenamefont {Linke}}]{Rad2022SurvivingTB}%
  \BibitemOpen
  \bibfield  {author} {\bibinfo {author} {\bibfnamefont {A.~I.}\ \bibnamefont {Rad}}, \bibinfo {author} {\bibfnamefont {A.}~\bibnamefont {Seif}},\ and\ \bibinfo {author} {\bibfnamefont {N.~M.}\ \bibnamefont {Linke}},\ }\bibfield  {title} {\bibinfo {title} {Surviving the barren plateau in variational quantum circuits with bayesian learning initialization}\ }(\bibinfo {year} {2022})\BibitemShut {NoStop}%
\bibitem [{\citenamefont {Volkoff}\ and\ \citenamefont {Coles}(2020)}]{Volkoff2020LargeGV}%
  \BibitemOpen
  \bibfield  {author} {\bibinfo {author} {\bibfnamefont {T.~J.}\ \bibnamefont {Volkoff}}\ and\ \bibinfo {author} {\bibfnamefont {P.~J.}\ \bibnamefont {Coles}},\ }\bibfield  {title} {\bibinfo {title} {Large gradients via correlation in random parameterized quantum circuits},\ }\href {https://api.semanticscholar.org/CorpusID:218869838} {\bibfield  {journal} {\bibinfo  {journal} {Quantum Science \& Technology}\ }\textbf {\bibinfo {volume} {6}} (\bibinfo {year} {2020})}\BibitemShut {NoStop}%
\bibitem [{\citenamefont {Ping}\ \emph {et~al.}(2013)\citenamefont {Ping}, \citenamefont {Li}, \citenamefont {Pan}, \citenamefont {Luo},\ and\ \citenamefont {Zhang}}]{Ping2013OptimalPO}%
  \BibitemOpen
  \bibfield  {author} {\bibinfo {author} {\bibfnamefont {Y.}~\bibnamefont {Ping}}, \bibinfo {author} {\bibfnamefont {H.}~\bibnamefont {Li}}, \bibinfo {author} {\bibfnamefont {X.}~\bibnamefont {Pan}}, \bibinfo {author} {\bibfnamefont {M.}~\bibnamefont {Luo}},\ and\ \bibinfo {author} {\bibfnamefont {Z.}~\bibnamefont {Zhang}},\ }\bibfield  {title} {\bibinfo {title} {Optimal purification of arbitrary quantum mixed states},\ }\href {https://api.semanticscholar.org/CorpusID:122400833} {\bibfield  {journal} {\bibinfo  {journal} {International Journal of Theoretical Physics}\ }\textbf {\bibinfo {volume} {52}},\ \bibinfo {pages} {4367} (\bibinfo {year} {2013})}\BibitemShut {NoStop}%
\bibitem [{Note4()}]{Note4}%
  \BibitemOpen
  \bibinfo {note} {We follow their idea, but we rewrite the derivation for clarity, as their paper has inconsistent notation.}\BibitemShut {Stop}%
\end{thebibliography}%

\onecolumngrid
\appendix

\section{Training State Unitary}
\label{app:training_state_unitary}

Let us write the quantum states of training data using the canonical basis: $\ket{\psi(\vb*{x}_i)}_\text{d}=\sum_\alpha c_{i}^\alpha\ket{\alpha}_\text{d}$.
We can write the training state in~\cref{eq:4} as
\begin{align}
    \rho_\text{train}=\dfrac{1}{N}\sum_{i=1}^N \ketbra{\psi(\boldsymbol{x}_i)} = \sum_{\alpha,\beta=0}^{2^{dn_x}-1} \left(\frac{1}{N}\sum_{i=1}^N c_i^\alpha (c_i^{\beta})^*\right)\ketbra{\alpha}{\beta}_\text{d}. \label{eq:rhotrain_expanded}
\end{align}

Now, let us denote by $U$ the tensor representation of the quantum circuit $U_\mathscr{D}$, which we will write using 4 indices:
\begin{align}
    U_\mathscr{D} = \sum_{\alpha,\gamma=0}^{2^{dn_x}-1}\sum_{\beta,\varepsilon=0}^{2^{n_a}}U^{\alpha\beta}_{\gamma\varepsilon}\ket{\alpha}_\text{d}\otimes\ket{\beta}_\text{a}\bra{\gamma}_\text{d}\otimes\bra{\varepsilon}_\text{a} = \sum_{\alpha\beta\gamma\epsilon}U^{\alpha\beta}_{\gamma\varepsilon}\ketbra{\alpha\beta}{\gamma\varepsilon},
\end{align}
where we will avoid writing tensor products, summation limits, and Hilbert space indices if there's no room for confusion.
Using this notation we can compute $\rho_\text{train}$ by partially tracing the auxiliary degrees of freedom of the state prepared with $U_\mathscr{D}$: 
\begin{align}
    \Tr_a(U_\mathscr{D}\ketbra{0}U^\dagger_{\mathscr{D}}) &= \sum_{j\alpha\beta\gamma\varepsilon\lambda\theta\varphi\phi} \mathbbm{1}_\text{d}\otimes\bra{j}_\text{a}U^{\alpha\beta}_{\gamma\varepsilon}\ketbra{\alpha\beta}{\gamma\varepsilon}\ketbra{00} (U^{\lambda\theta}_{\varphi\phi})^* \ketbra{\varphi\phi}{\lambda\theta} \mathbbm{1}_\text{d}\otimes\ket{j}_\text{a}\\
    &= \sum_{j\alpha\beta\gamma\varepsilon\lambda\theta\varphi\phi} U^{\alpha\beta}_{\gamma\varepsilon} (U^{\lambda\theta}_{\varphi\phi})^*\delta_{j\beta}\delta_{\gamma0}\delta_{\varepsilon0}\delta_{\varphi0}\delta_{\phi0}\delta_{\theta j}\ketbra{\alpha}{\lambda}_\text{d}\\
    &=\sum_{j=0}^{2^{n_a}-1}\sum_{\alpha,\lambda=0}^{2^{dn_x}-1} U^{\alpha j}_{00}(U^{\lambda j}_{00})^* \ketbra{\alpha}{\lambda}_\text{d}\\
    &= \sum_{\alpha,\beta=0}^{2^{dn_x}-1}\left(\sum_{j=0}^{2^{n_a}-1}U^{\alpha j}_{00}(U^{\beta j}_{00})^* \right)\ketbra{\alpha}{\beta}_\text{d}.
\end{align}
By comparison with~\cref{eq:rhotrain_expanded}, we obtain
 \begin{align}
    \frac{1}{N}\sum_{i=1}^N c_i^\alpha (c_i^{\beta})^* = \sum_{j=0}^{2^{n_a}-1}U^{(\alpha j)}_{(00)}\left(U^{(\beta j)}_{(00)}\right)^*,\label{eq:condition_on_U_D}
    \end{align}
which must be satisfied for every $\alpha$ and $\beta$.
Since the columns of a unitary matrix form an orthonormal basis, one must have that $\sum_{\alpha j}|U_{00}^{\alpha j}|^2 = 1$.
A naïve option to satisfy such requirement is to define $U_{00}^{\alpha j} = \frac{1}{\sqrt{N}}c_i^\alpha$, where we can match the indices $i$ and $j$ by using $n_a = \log_2(N)$ auxiliary qubits.
$U$ can be written as a matrix as well, by combining the superscript indices into one, and the subscript indices into another one. The norm of the first column is $\sum_{\alpha j}|U_{00}^{\alpha j}|^2 = \sum_{\alpha i} \frac{1}{N}|c_i^\alpha|^2=\sum_{i}\frac{1}{N} = 1$.
The rest of the columns can be built using the Gram-Schmidt process.

An optimal solution, however, can be reached for $n_a=dn_x$, which is independent from the number of training data points $N$; for this, let us follow~\citet{Ping2013OptimalPO}\footnote{We follow their idea, but we rewrite the derivation for clarity, as their paper has inconsistent notation.}.
In general, \cref{eq:rhotrain_expanded} admits the factorisation $\rho_{\text{train}} = V\Lambda V^\dagger$, where $V$ is the unitary $V=\sum_{\alpha,\beta=0}^{m_\text{d}-1}V_{\alpha\beta}\ketbra{\alpha}{\beta}_\text{d}$ and $\Lambda$ is a diagonal matrix $\text{diag}(\lambda_0,\ldots,\lambda_{m_\text{d}-1})$, where $m_\text{d}:=2^{dn_x}$ is defined to shorten the notation. 
We need to build a state of the form $\ket{\Phi}=\sum_{\alpha\beta}V_{\alpha\beta}\sqrt{\lambda_\beta}\ket{\alpha\beta}$, which is a purification of $\rho_\text{train}$:
\begin{align}
    \Tr_\text{a}(\ketbra{\Phi}) &= \sum_{j\alpha\beta\gamma\varepsilon} \mathbbm{1}_\text{d}\otimes\bra{j}_\text{a}V_{\alpha\beta}\sqrt{\lambda_\beta}\ketbra{\alpha\beta}{\gamma\varepsilon}\sqrt{\lambda^*_\varepsilon}V^*_{\gamma\varepsilon}\mathbbm{1}_\text{d}\otimes\ket{j}_\text{a}\\
    &= \sum_{j\alpha\beta\gamma\varepsilon}V_{\alpha\beta}V^*_{\gamma\varepsilon}\sqrt{\lambda_\beta\lambda_\varepsilon^*}\delta_{j\beta}\delta_{\varepsilon j} \ketbra{\alpha}{\gamma} = \sum_{j\alpha\gamma}V_{\alpha j} V^*_{\gamma j}\lambda_j\ketbra{\alpha}{\gamma}= V\Lambda V^\dagger.
\end{align}
Preparing such state can be done using the aforementioned Gram-Schmidt process.
Note that if $\rho_\text{train}$ is not full-rank, then, some $\lambda_\beta$ values would be zero, effectively making $n_a=\log_2 r$, where $r$ is the rank.

\section{Fitness and Loss Functions}
\label{app:fitness_and_loss}

Encoding variational quantum circuits in bitstrings enables the application of genetic and memetic algorithms to find the best circuit for solving the kernel approximation problem. 
The genetic algorithm explores the space of bitstrings representing variational quantum circuit architectures with fixed angles, thereby optimising the fitness function:

\begin{align}
    \min_{\vb*{b}}\sum_{(x,x')\sim U^2(a,b)}\left(|k(x,x')|^2  - \abs{\bra{0}_\text{f}\tilde{U}^\dagger_\star(\vb*{b},x)\tilde{U}_\star(\vb*{b},x')\ket{0}_\text{f}}^2\right)^2.
    \label{eq:genetic_fitness}
\end{align}

The memetic algorithm incorporates local optimisation through a gradient descent routine that optimises the decoded angles of each individual, effectively converting them into agents. This allows the algorithm to explore both the bitstring space of architectures and the parameter space of angles, applying the fitness function of \cref{eq:single_feature_cost_fn}:

\begin{equation*}
    \min_{\vb*{b},\vb*{\theta}_{\vb*{b}}}\sum_{(x,x')\sim U^2(a,b)}\left(|k(x,x')|^2  - \abs{\bra{0}_\text{f}\tilde{U}^\dagger_\star(\vb*{b},\vb*{\theta}_{\vb*{b}},x)\tilde{U}_\star(\vb*{b},\vb*{\theta}_{\vb*{b}},x')\ket{0}_\text{f}}^2\right)^2.
\end{equation*}

Moreover, it is also possible to refrain from exploring the space of variational quantum circuit architectures and instead optimise the angles of a fixed architecture such as the HEA via the loss function:

\begin{equation}
    \min_{\vb*{\theta}}\sum_{(x,x')\sim U^2(a,b)}\left(|k(x,x')|^2  - \abs{\bra{0}_\text{f}\tilde{U}^\dagger_\star(\vb*{\theta},x)\tilde{U}_\star(\vb*{\theta},x')\ket{0}_\text{f}}^2\right)^2.
    \label{eq:hea_loss}
\end{equation}

\section{Experimental Parameters}
\label{app:experimental_parameters} 

In \cref{sec:qfm_results}, the kernel is approximated using genetic, memetic, and gradient descent algorithms, optimising Eqs. \ref{eq:genetic_fitness}, \ref{eq:single_feature_cost_fn}, and \ref{eq:hea_loss}, respectively.
To ensure a fair comparison between the three methods, specific experimental parameters were chosen for each to ensure accurate inference of the kernel from a set of unknown pairs and proper convergence of the fitness or loss function values. 
\Cref{tab:parameters} displays the experimental parameters utilised in comparing the three methods.
Additionally, in all cases, the optimisation routines were configured with a Gaussian kernel bandwidth of $0.1$, using $10^4$ training pairs within the interval $[-3,3]$, i.e., $(x,x')\sim U^2[-3,3]$.

\begin{table}[]
    \centering
    \begin{tabular}{cccc}
    \toprule
        & Genetic & Memetic & Gradient Descent \\
    \midrule
        Generations & 30 & 30 & n/a \\
        Individuals & 15 & 15 & n/a \\
        Epochs & n/a & 2000 & 2000 \\
        Learning Rate & n/a & 0.2 & 0.2 \\
        KLD Seeds & 50 & 50 & 50 \\
    \bottomrule
    \end{tabular}
    \caption{Experimental parameters used in the comparison of the three different methods approximating the Gaussian kernel.}
    \label{tab:parameters}
\end{table}

\section{Data generation for KLD}
\label{app:data_KLD}

\begin{figure}[H]
    \centering
    \includegraphics[scale = 0.93]{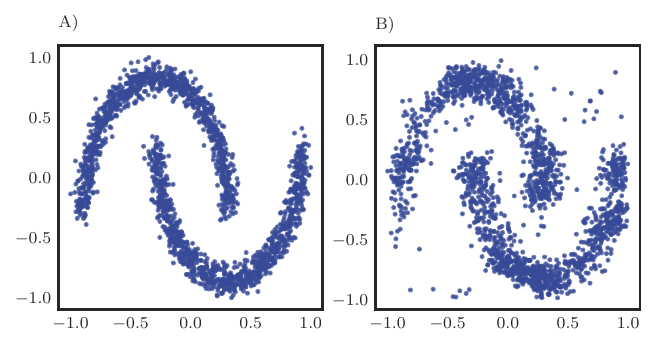}
    \caption{Sample sets used to estimate the Kullback-Leibler divergence (KLD) between the true density and the density estimated by the DMKDE circuit. A) Samples of the training set $\mathcal{X}$ (generated from the true density $t(\boldsymbol{x})$). B) One of the sets of samples $\mathcal{X'}$ generated by the DMKDE circuit (3 QFM qubits and 5 training circuit layers) from the estimated density $p(\boldsymbol{x})$. By averaging the KLD value estimates between the training data and different datasets generated through the DMKDE circuit, we can provide a measure of the circuit's performance in density estimation.}
    \label{fig:KLD}
\end{figure}

To evaluate the performance of the DMKDE circuit, we follow the implementation of \citet{PrezCruz2008KullbackLeiblerDE} to estimate the Kullback-Leibler divergence (KLD) for continuous distributions. 
Given $n$ independent and identically distributed (i.i.d) samples $\mathcal{X}=\left\{\boldsymbol{x}_i\right\}_{j=1}^n$ from the true density $t(\boldsymbol{x})$ and $m$ i.i.d samples $\mathcal{X}'=\left\{x'_j\right\}_{j=1}^m$ from the predicted density $p(\boldsymbol{x})$, we can estimate the KLD measure from a k-nearest-neighbour density estimate as

\begin{equation}
    \hat{D}(T||P)=\dfrac{d}{n}\sum_{i=1}^n\log\dfrac{r_k(\boldsymbol{x_i})}{s_k(\boldsymbol{x}_i)}+\log\dfrac{m}{n-1},
\end{equation}

where $r_k(\boldsymbol{x}_i)$ and $s_k(\boldsymbol{x}_i)$ are, respectively, the Euclidean distance to the $k^{th}$ nearest-neighbour of $\boldsymbol{x}_i$ in $\mathcal{X}\setminus\{\boldsymbol{x}_i\}$ and $\mathcal{X}'$, and $d$ is the number of data dimensions.

It is therefore necessary to generate samples from the density estimated by DMKDE ($p(\boldsymbol{x})$). To achieve this, we adopt the strategy of generating data uniformly over the interval of the original data and then selecting only points with a probability determined by the value of the density estimate, as shown in \cref{fig:KLD}. 
This enables us to generate a set of samples from the density estimated by DMKDE ($p(\boldsymbol{x})$) and assess the performance of the circuit's density estimation by estimating the Kullback-Leibler divergence (KLD) value between the real density $t(\boldsymbol{x})$ and the estimated density $p(\boldsymbol{x})$. 
However, it is important to note that the generation of data from the DMKDE estimation is a stochastic process. 
Therefore, to ensure accuracy, it is necessary to use different generation seeds and average the estimated KLD values. 
Additionally, standard deviation is also taken into account.

\end{document}